\documentclass[%
    rmp,
    amsmath,
    amssymb,
    noeprint,
    titlepage
]{revtex4-2}

\usepackage{graphicx}
\usepackage[%
    colorlinks=true,
    allcolors=blue
]{hyperref}
\usepackage{siunitx}
\usepackage[%
    breakable,
    skins
]{tcolorbox}
\usepackage{bm}

\DeclareSIUnit{\pc}{\text{pc}}
\DeclareSIUnit{\yr}{\text{yr}}
\DeclareSIUnit{\gauss}{\text{G}}
\DeclareSIUnit{\dyn}{\text{dyn}}
\DeclareSIUnit{\day}{\text{d}}

\newtcolorbox[auto counter]{tbox}[2][]{%
    colframe=orange!75!black,
    fonttitle=\bfseries,
    before upper={\parindent15pt},
    breakable,
    enhanced,
    title=Box~\thetcbcounter: #2,#1
}

\begin{document}

\title{Gravitational waves from neutron-star mountains}

\author{Fabian Gittins}
\email{f.w.r.gittins@soton.ac.uk}
\affiliation{Mathematical Sciences and STAG Research Centre, University of Southampton, Southampton SO17 1BJ, United Kingdom}

\date{\today}

\begin{abstract}
Rotating neutron stars that support long-lived, non-axisymmetric deformations known as mountains have long been considered potential sources of gravitational radiation. However, the amplitude from such a source is very weak and current gravitational-wave interferometers have yet to witness such a signal. The lack of detections has provided upper limits on the size of the involved deformations, which are continually being constrained. With expected improvements in detector sensitivities and analysis techniques, there is good reason to anticipate an observation in the future. This review concerns the current state of the theory of neutron-star mountains. These exotic objects host the extreme regimes of modern physics, which are related to how they sustain mountains. We summarise various mechanisms that may give rise to asymmetries, including crustal strains built up during the evolutionary history of the neutron star, the magnetic field distorting the star's shape and accretion episodes gradually constructing a mountain. Moving beyond the simple rotating model, we also discuss how precession affects the dynamics and modifies the gravitational-wave signal. We describe the prospects for detection and the challenges moving forward.
\end{abstract}

\maketitle

\section{Introduction}

We live in a fortuitous time when it comes to gravitational-wave astronomy. Since the inaugural detection of the black-hole binary GW150914 \cite{2016PhRvL.116f1102A}, sensitive, ground-based, gravitational-wave instruments---LIGO \cite{2015CQGra..32g4001L} and Virgo \cite{2015CQGra..32b4001A}---have witnessed approximately $100$ compact-binary mergers in the first three observing runs \cite{2019PhRvX...9c1040A,2021PhRvX..11b1053A,2021arXiv210801045T,2021arXiv211103606T}. Indeed, with no corresponding electromagnetic counterpart to the black-hole events (that we have seen, at least), gravitational-wave interferometers provide astronomers a new lens to observe the optically dark sector of the Universe. In addition to black holes, detectors have also seen coalescences involving neutron stars, which include neutron star-neutron star \cite{2017PhRvL.119p1101A,2020ApJ...892L...3A} and neutron star-black hole mergers \cite{2021ApJ...915L...5A}. Among these detections is the spectacular event GW170817 \cite{2017PhRvL.119p1101A}, which was accompanied by a host of electromagnetic signals across the spectrum \cite{2017ApJ...848L..12A}, thereby beginning a new era of multimessenger astronomy.

Given this remarkable success, it would be quite easy to take such measurements from far-away, cataclysmic events for granted. Naturally, it would be remiss to forget the shear scale of the original problem: detecting strains at least of the order of $10^{-21}$ \cite{2000CQGra..17.2441S}. For the kilometre-length arms of LIGO and Virgo, this corresponds to measuring displacements the size of a fraction of a proton. In a feat of experimental and data-analytical triumph, this level of precision has been reached, enabling accurate measurements of the gravitational-wave strain. In tandem with this achievement, theory has played a vital role in developing waveforms to match the strain measurements. However, we surely cannot be blamed for dreaming of more. And with third-generation detectors on the horizon---The Einstein Telescope \cite{2010CQGra..27s4002P} and Cosmic Explorer \cite{2019BAAS...51g..35R}---with vastly improved design sensitivities, there is a good chance that we will see more.

There are good reasons to be particularly excited by the prospect of further gravitational-wave signals from neutron stars. Perhaps the most compelling reason lies in the simple fact that they are material bodies, in contrast to vacuum black holes. Although they are far from typical stars---due to their compactness, neutron-star interiors reach densities in excess of nuclear saturation. For this reason, they likely hold the key to understanding ultra-dense nuclear matter; in particular, the thermodynamic \textit{equation of state} \cite{2001ApJ...550..426L,2007PhR...442..109L,2007ASSL..326.....H,2012ARNPS..62..485L,2016ARA&A..54..401O}. Indeed, we have begun probing this information since the high signal-to-noise ratio event GW170817 provided an upper limit on the tidal deformabilities of the involved neutron stars, which in turn constrains the equation of state \cite{2017PhRvL.119p1101A,2018PhRvL.121p1101A,2019PhRvX...9a1001A}. There are numerous other aspects of neutron-star physics that make them veritable astrophysical laboratories, including their strong magnetic fields, solid crusts and high spin frequencies to name but a few, and gravitational waves present the very real possibility of probing much of this physics, in tandem with electromagnetic observations.

At the risk of looking a gift horse in the mouth, we are yet to witness gravitational radiation from the simple case of a deformed, rotating neutron star. But why should we expect neutron stars to support such deformations in the first place? The answer is quite simple, even though the precise details are not. The formation history of a neutron star is expected to be quite complex. They begin their lives as the remnant of a supernova. Newly born neutron stars will be hot and rapidly rotating, as they retain some of the angular momentum of the progenitor star. As the star cools, its crust will solidify, since it becomes energetically favourable at low densities to form a crystal lattice. The neutron star may, at some point in its lifetime, exhibit glitches and accrete from a companion star. Many of these processes will build up strain in the crust and cause the star to change shape. It is therefore inevitable that the typical neutron star, much like planets with a solid component, will be deformed in a non-axisymmetric fashion and emit gravitational waves. In addition to strain in the crust, there are other ways in which an asymmetry can develop, which we will explore in this review.

In this review, our focus will be on the physics of deformed neutron stars. This review is not intended to be exhaustive by any means and the objective is provide an accessible overview of this research area. Throughout this article, references are provided to important review articles, textbooks and selected papers, which will be useful to readers less familiar with the topics discussed. For related reviews on deformed neutron stars, see \citet{2002CQGra..19.1255J}, \citet{2009ASSL..357..651P}, \citet{2015PASA...32...34L} and \citet{2018ASSL..457..673G}. We will primarily consider relatively long-lived perturbations known as \textit{mountains}. These are separate to oscillation modes, which may become excited and radiate gravitational waves. Since mountains are long lived, they are promising candidates for continuous gravitational-wave detection. There are complementary reviews on this subject, which the reader is referred to for more information \cite{2012arXiv1201.3176P,2019Univ....5..217S,2021Univ....7..474T,2022Galax..10...72P,2023NatAs...7.1160H,2023APh...15302880W,2023LRR....26....3R}.

This article is structured as follows. We begin in Sec.~\ref{sec:Motivation} with the motivation for why neutron-star mountains are interesting from a gravitational-wave perspective. With this motivation in hand, we move on to discuss how neutron stars can develop such asymmetries. In Sec.~\ref{sec:Crust}, we review the involvement of the crust in supporting mountains, particularly with regards to how large the deformations can become before the crust fractures. Next, in Sec.~\ref{sec:Magnetic}, we consider how the magnetic field can distort the star into a quadrupolar shape and discuss estimated mountain sizes. In Sec.~\ref{sec:Accretion}, we discuss how accretion can give rise to mountains in low-mass X-ray binaries. This scenario includes mountains sourced by temperature-sensitive reactions (Sec.~\ref{sec:Thermal}) and those confined by the magnetic field (Sec.~\ref{sec:Confinement}). Following these mechanisms, in Sec.~\ref{sec:Precession}, we consider how precession modifies the picture. We summarise the prospects for continuous gravitational-wave detection of neutron-star mountains in Sec.~\ref{sec:Prospects}, highlighting the challenges that lie ahead. We summarise and conclude in Sec.~\ref{sec:Summary}.

In this article, we use the metric signature $(-, +, +, +)$. The Einstein summation convention is adopted, where repeated indices denote a summation. We use early Latin characters $a, b, c, \ldots$ for spacetime indices and later characters $i, j, k, \ldots$ for spatial indices. We reserve the characters $(l, m)$ for spherical harmonics. We will also encounter the following physical constants: Newton's gravitational constant $G$, the speed of light in a vacuum $c$ and Boltzmann's constant $k_\text{B}$. So, without further ado, let us begin.

\section{Motivation}
\label{sec:Motivation}

Gravitational waves are a remarkable prediction of Einstein's general theory of relativity. Since the first detection of gravitational waves in 2015, we have a new way to observe the Universe. In this section, we will provide an overview of how gravitational waves are produced in general relativity. We will see that they are emitted when a massive source has a time-varying mass quadrupole moment. We will show how a rotating neutron star that is deformed non-axisymmetrically emits gravitational radiation. Since we detect many spinning pulsars using conventional electromagnetic telescopes, we briefly discuss the tentative evidence for gravitational waves from such observations.

\subsection{Gravitational-wave theory}
\label{sec:GravitationalWaves}

In general relativity, accelerating masses generate gravitational radiation.%
\footnote{This is analogous to how accelerating charges produce electromagnetic radiation. Indeed, there exist many analogies between general relativity and electromagnetism, which has been the subject of some study and termed \textit{gravito-electromagnetism} \cite{1998CQGra..15..705M,2003gr.qc....11030M}.}
It is a standard textbook exercise to show that metric perturbations $h_{a b}$ in spacetime satisfy a wave equation \cite{1973grav.book.....M,2003gieg.book.....H,Maggiore2008,2014grav.book.....P,2019gwa..book.....A,Schutz2022}
\begin{equation}
    \Box \bar{h}_{a b} = - \frac{16 \pi G}{c^4} T_{a b},
    \label{eq:Wave}
\end{equation}
where $\Box = \partial_a \partial^a$ is the (flat) d'Alembertian, $\bar{h}_{a b} = h_{a b} - \eta_{a b} h_c^c / 2$ is the trace-reversed metric perturbation, $\eta_{a b}$ is the Minkowski metric and $T_{a b}$ is the contribution to the stress-energy tensor of the source that enters at linear order in the perturbations (we will soon see that it is the part of the stress-energy tensor due to asymmetric matter motion). In deriving Eq.~\eqref{eq:Wave}, it has been assumed that the perturbation exists on top of a Minkowskian spacetime and the coordinates have been chosen such that the metric perturbation satisfies the harmonic gauge $\partial^b \bar{h}_{a b} = 0$.%
\footnote{The assumption that the background is flat is particularly restrictive and is inapplicable to many situations of astrophysical interest. Fortunately, this assumption can be relaxed by averaging over several wavelengths, but the calculation is quite involved. For our pedagogical purposes, it will be sufficient to assume a flat background.}
Equation~\eqref{eq:Wave} shows that general relativity permits small perturbations that travel at the speed of light. Thus, we call the solutions gravitational waves.

The solution to Eq.~\eqref{eq:Wave} can be obtained using a retarded Green's function,
\begin{equation}
    \bar{h}_{a b}(t, x^i) = \frac{4 G}{c^4} \int \frac{T_{a b}(t' = t - |x^i - x'^i| / c, x'^i)}{|x^i - x'^i|} \, dV',
\end{equation}
where the strain is calculated at a spacetime point $(t, x^i)$ in the harmonic gauge and the integration is taken over the volume of the source, using coordinates $(t', x'^i)$, with volume element $dV'$. In order to understand the features, it is convenient to consider a region of space far from a weak-gravity source. In this context, with the help of the conservation equation $\partial_b T^{a b} = 0$, we find
\begin{equation}
    \bar{h}_{i j}(t, x^i) = \frac{2 G}{c^6 d} \frac{d^2}{dt^2} \int T^{t t}(t' = t - d / c, x'^i) x'_i x'_j \, dV' + \mathcal{O}(d^{-2}) = \frac{4 G}{c^4 d} \frac{d^2 M_{i j}}{dt^2}(t - d / c) + \mathcal{O}(d^{-2}, c^{-6}),
\end{equation}
where $d = |x^i - x'^i|$ is the distance to the source and we have identified its (Newtonian) \textit{mass quadrupole moment tensor} $M_{i j} = \int \rho x_i x_j \, dV$ from $T^{t t} = \rho c^2 + \mathcal{O}(1)$ with $\rho$ the mass density. This is formally a post-Newtonian expansion of general relativity, which is appropriate in a region where the source's gravity is weak \cite{1973grav.book.....M,2007LRR....10....2F,2014grav.book.....P,2014LRR....17....2B,2020RPPh...83g5901L}. However, we expect, and indeed know, that the strongest gravitational-wave emitters are hardly weak-gravity sources. One would assume (quite reasonably) that the post-Newtonian approximation would be fairly ineffective at describing such compact objects. It turns out that this is not the case, and the post-Newtonian expansion is very useful and fairly indispensable for describing these kind of systems, even during the late inspiral and merger of two black holes \cite{2011PNAS..108.5938W}. In practice, the mass quadrupole moment that one reads off from the expansion corresponds to the relativistic quantity \cite{1980RvMP...52..299T}. For a summary of the multipole moments, see Box~\ref{box:Multipole}.

\begin{tbox}[label=box:Multipole]{Multipole moments}
Spherical symmetry is a convenient simplification, but it rarely describes celestial bodies in Nature. For neutron stars, rotation and strains built up in the crust will cause the star to deviate from sphericity. When these deviations are relatively small, multipole expansions are a powerful way to describe the star's shape.

We consider an observer at position $x^i$ outside a Newtonian matter source with mass density $\rho$. The observer will measure the exterior gravitational field of the source $\Phi$---which is governed by Poisson's equation $\nabla^2 \Phi = 4 \pi G \rho$---as
\begin{equation*}
    \Phi(t, x^i) = - G \int \frac{\rho(t, x'^i)}{|x^i - x'^i|} \, dV'.
\end{equation*}
The factor of $|x^i - x'^i|^{-1}$ can be expressed in terms of the spherical harmonics $Y_l^m(\theta, \phi)$, so that
\begin{equation*}
    \Phi(t, r, \theta, \phi) = - G \sum_{l = 0}^\infty \sum_{m = - l}^l \frac{4 \pi}{2 l + 1} Q_{l m}(t) \frac{Y_l^m(\theta, \phi)}{r^{l + 1}},
\end{equation*}
where $r = |x^i|$ and we have introduced the \textit{multipole moments} 
\begin{equation*}
    Q_{l m}(t) = \int \rho(t, x^i) r^l Y_l^{m*}(\theta, \phi) \, dV,
\end{equation*}
with $^*$ denoting a complex conjugate. We may decompose a generic scalar function using spherical harmonics, 
\begin{equation*}
    \rho(t, r, \theta, \phi) = \sum_{l = 0}^\infty \sum_{m = -l}^l \rho_{l m}(t, r) Y_l^m(\theta, \phi).
\end{equation*}
Inserting this decomposition into $Q_{l m}$ leads to
\begin{equation*}
    Q_{l m}(t) = \int \rho_{l m}(t, r) r^{l + 2} \, dr.
\end{equation*}
Thus, each harmonic of the mass density $\rho$ sources a multipole moment $Q_{l m}$.

There exists an alternative (but equivalent) formulation of the multipole moments involving symmetric, trace-free tensors. Instead of expanding $|x^i - x'^i|^{-1}$ with spherical harmonics, one can instead use a Taylor series. Thus, in a Cartesian basis, 
\begin{equation*}
    \Phi = - G \left[ \frac{M}{r} + \frac{p^i n_i}{r^2} + \frac{3}{2} \frac{Q^{i j}}{r^3} \left( n_i n_j - \frac{1}{3} \delta_{i j} \right) \right] + \mathcal{O}(r^{-4}),
\end{equation*}
where $n_i = \partial_i r$ is a unit vector and we identify the monopole $M = \int \rho \, d^3 x$, dipole $p^i = \int \rho x^i \, d^3 x$ and trace-free quadrupole [\textit{cf.}, Eq.~\eqref{eq:Quadrupole}]
\begin{equation*}
    Q^{i j} = \int \rho \left( x^i x^j - \frac{1}{3} r^2 \delta^{i j} \right) \, dV.
\end{equation*}
We are always free to choose a coordinate system where $p^i = 0$. This is the centre-of-mass frame. For a discussion of the multipole moments in general relativity, where mass multipoles are joined by the relativistic current multipoles, see \citet{1980RvMP...52..299T}.
\end{tbox}

It turns out that we have residual freedom to choose a specific set of coordinates within the harmonic gauge. It is common to choose the transverse-traceless gauge denoted by $h_{i j}^\text{TT}$,%
\footnote{The transverse-traceless gauge picks out a particular frame where the metric perturbation is purely spatial $h_{t i}^\text{TT} = 0$ and trace-free $h_i^{i \, \text{TT}} = 0$. The harmonic gauge condition implies that such a perturbation is transverse, $\partial^j h_{i j}^\text{TT} = 0$.}
where we now have 
\begin{equation}
    h_{i j}^\text{TT}(t, x^i) = \frac{4 G}{c^4 d} \frac{d^2 Q_{i j}}{dt^2} (t - d / c) + \mathcal{O}(d^{-2}, c^{-6})
    \label{eq:StrainQuadrupole}
\end{equation}
with the \textit{traceless} mass quadrupole moment tensor
\begin{equation}
    Q_{i j} = \int \rho \left( x_i x_j - \frac{1}{3} r^2 \delta_{i j} \right) dV,
    \label{eq:Quadrupole}
\end{equation}
where we have adopted a Cartesian basis. Here, we have Einstein's famous \textit{quadrupole formula}~\eqref{eq:StrainQuadrupole}, which serves as the basis of many useful gravitational-wave estimates \cite{1918SPAW.......154E,1971ctf..book.....L}. We observe that gravitational radiation (i) is sourced by accelerating massive bodies, (ii) falls off as $1 / d$ and (iii) is quadrupolar at leading post-Newtonian order. The absence of monopole radiation is a consequence of Birkoff's theorem \cite{1921ArMAF..15...18J,1923rmp..book.....B}, whereas dipole radiation does not exist due the lack of negative charges in gravity (in stark contrast to electromagnetic waves).

An important step in the development of the theory was the demonstration that gravitational waves carry energy. This component of the theory was the subject of much controversy in its development \cite{2007tste.book.....K}. A particular issue was the fact that the effective stress-energy tensor of the waves $t_{a b}$ was gauge-dependent; one could simply choose an inertial frame where the radiation vanishes. It was eventually shown that the physical effect of the waves manifests over several cycles \cite{PhysRev.166.1272},
\begin{equation}
    \langle t_{a b} \rangle = \frac{c^4}{32 \pi G} \langle \partial_a h_{i j}^\text{TT} \partial_b h^{i j \, \text{TT}} \rangle,
    \label{eq:IsaacsonTensor}
\end{equation}
where $\langle \ldots \rangle$ denotes an average over wavelengths. This form, the Isaacson stress-energy tensor~\eqref{eq:IsaacsonTensor}, is gauge invariant, as it must be in order to be physically relevant. The Isaacson stress-energy tensor~\eqref{eq:IsaacsonTensor} describes the energy the gravitational radiation carries away from the source, which can be combined with the quadrupole formula~\eqref{eq:StrainQuadrupole} to obtain the total gravitational-wave luminosity
\begin{equation}
    \mathcal{F} = \frac{dE_\text{GW}}{dt} = \frac{G}{5 c^5} \left\langle \frac{d^3 Q_{i j}}{dt^3} \frac{d^3 Q^{i j}}{dt^3} \right\rangle + \mathcal{O}(c^{-7}).
    \label{eq:Reaction}
\end{equation}
This effect is known as the \textit{gravitational-radiation reaction} on the source. Gravitational waves carry energy away from the source at a rate given by Eq.~\eqref{eq:Reaction}.

In summary, provided the source has a time-varying quadrupole moment (or higher multipole), then it will emit gravitational radiation. Compact objects are the most promising candidates, since they strongly disturb spacetime. Of course, we have seen radiation from compact binaries, but the focus of this review is on a different source.

\subsection{Rotating, deformed neutron stars}
\label{sec:Rotating}

An isolated body must spin and be deformed away from rotational axisymmetry in order to gravitationally radiate.%
\footnote{Although a rotating body will inherit a natural quadrupolar deviation from (spherical) symmetry due to the centrifugal force, this will not be sufficient to emit gravitational waves. The wave solution~\eqref{eq:StrainQuadrupole} shows that the quadrupole moment of the body must vary in time with respect to its second time derivative.}
Neutron stars possess solid crusts close to the surface, which enable them to support long-lived, non-axisymmetric perturbations. We will examine how a rotating body emits gravitational waves.

In this review, we will treat the star in an entirely Newtonian setting. We do this to illustrate the physics, since there is still substantial uncertainty surrounding the different ingredients required to model real neutron stars. However, caution should be exercised. Although it is tempting to consider general relativity as a ``correction'' to the Newtonian theory (where results are typically altered at the 15--20\% level), for compact objects, gravity is a leading-order effect. The stellar-structure equations provide an example of this. It is well known that incorporating realistic supranuclear equations of state into a Newtonian description of the structure provide spurious masses and radii that differ greatly from the relativistic results. In order to construct realistic neutron-star models, we will need to bring all the physics together within full, gory general relativity. At present, we are some way from the final description, but this should be our ultimate goal.

\begin{figure}[h]
    \includegraphics[width=0.55\textwidth]{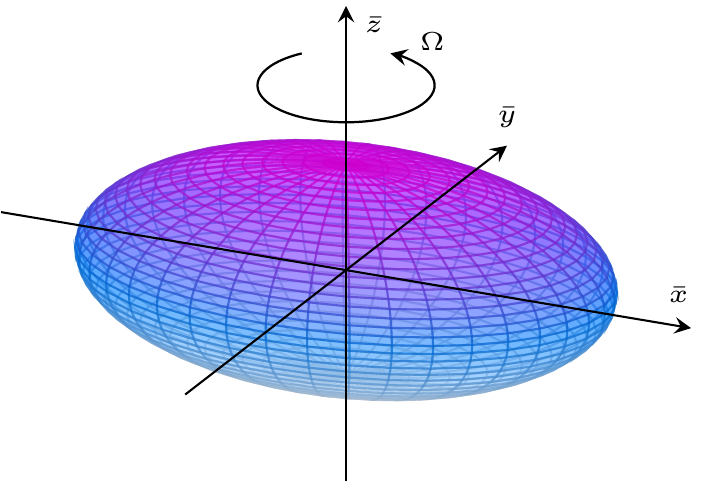}
    \caption{\label{fig:DeformedStar}%
    An illustration of a non-axisymmetrically deformed star rotating with angular velocity $\Omega$. As the star rotates, its body frame $(\bar{x}, \bar{y}, \bar{z})$ co-rotates with it. The star's shape is characterised by its principal moments of inertia $(I_1, I_2, I_3)$, defined as the moment $I_1$ about its $\bar{x}$-axis, $I_2$ about its $\bar{y}$-axis and $I_3$ about its $\bar{z}$-axis.
    }
\end{figure}

We begin by erecting an inertial, Cartesian coordinate system $(x, y, z)$. In this coordinate system, we place a uniformly rotating star with angular velocity $\Omega$ about the $z$-axis, where the origin coincides with the centre of mass. The star carries with it a co-rotating coordinate system $(\bar{x}, \bar{y}, \bar{z})$ known as the body frame, shown in Fig.~\ref{fig:DeformedStar}, where we align $\bar{z} = z$ for simplicity. We endow the star with a non-trivial mass quadrupole and treat it as rigid. For the time being, we are not concerned with the stellar interior.

We consider the star's moment of inertia tensor
\begin{equation}
    I_{i j} = \int \rho (r^2 \delta_{i j} - x_i x_j) \, dV,
\end{equation}
which can be straightforwardly related to the quadrupole tensor~\eqref{eq:Quadrupole}. In the body frame, the star has principal moments of inertia $(I_1, I_2, I_3)$ defined in the body frame by 
\begin{equation}
    [I_{\bar{i} \bar{j}}] = 
    \begin{bmatrix}
        I_1 & 0 & 0\\
        0 & I_2 & 0\\
        0 & 0 & I_3
    \end{bmatrix}.
\end{equation}
The star is non-axisymmetric if at least one of $I_1$ and $I_2$ differs from $I_3$. We can simply transform to the inertial frame with a rotation in the $\bar{x}-\bar{y}$ plane to obtain
\begin{equation}
    [I_{i j}] =
    \begin{bmatrix}
        \frac{1}{2}  (I_1 + I_2) + \frac{1}{2} (I_1 - I_2) \cos (2 \phi) & \frac{1}{2} (I_1 - I_2) \sin (2 \phi) & 0\\
        \frac{1}{2} (I_1 - I_2) \sin (2 \phi) & \frac{1}{2}  (I_1 + I_2) - \frac{1}{2} (I_1 - I_2) \cos (2 \phi) & 0\\
        0 & 0 & I_3
    \end{bmatrix},
\end{equation}
where $\phi$ is the angle between the two frames. Since the rotation is uniform, we are free orientate the frames such that $\phi = \Omega t$. Here, we see that a rotating, quadrupolar source will radiate gravitational waves at twice the rotation rate $2 \dot{\phi}$, where a dot denotes a time derivative.%
\footnote{This is also true for binaries, where the waves are emitted at twice the orbital frequency. However, if the radiation is generated by an oscillation mode of the star, then the gravitational waves will be emitted at the frequency of the mode. Additionally, we will see later that when precession is involved gravitational waves are also emitted at the spin frequency.}
\footnote{Although the star will experience centrifugal deformation due to the rotation, this will be a constant contribution to the moment of inertia.}
Gravitational waves will also be emitted at a second frequency of $\dot{\phi}$ under the influence of mechanisms such as precession (which we discuss later in Sec.~\ref{sec:Precession}), magnetic fields \cite{1996A&A...312..675B} and pinned superfluidity \cite{2010MNRAS.402.2503J}.

We are now in a position to obtain some useful estimates. For these calculations, we will rely on the post-Newtonian formulae presented in Sec.~\ref{sec:GravitationalWaves}. Recall that this formalism applies for weakly relativistic sources and so, for the neutron stars that we are interested in, the post-Newtonian approximation will be used for order-of-magnitude estimates. By Eq.~\eqref{eq:StrainQuadrupole}, we introduce the strain amplitude 
\begin{equation}
    h_0 = \frac{4 G}{c^4} \frac{\epsilon I_3 \Omega^2}{d} \approx 10^{-25} \left( \frac{\qty{10}{\kilo\pc}}{d} \right) \left( \frac{\epsilon}{10^{-6}} \right) \left( \frac{I_3}{\qty{e45}{\g\cm\squared}} \right) \left( \frac{\nu}{\qty{500}{\hertz}} \right)^2,
    \label{eq:StrainAmplitude}
\end{equation}
where $\nu = \Omega / (2 \pi)$ is the star's spin frequency and we have defined the \textit{ellipticity} of the star
\begin{equation}
    \epsilon \equiv \frac{I_2 - I_1}{I_3},
    \label{eq:Ellipticity}
\end{equation}
which is a dimensionless measure of the star's quadrupolar deviation from symmetry about the $\bar{z}$-axis. We see from Eq.~\eqref{eq:StrainAmplitude} that, even for a reasonably close and rapidly rotating source, the gravitational-wave strain is extremely weak. The nearby Crab pulsar PSR B0531+21 is useful to consider. It is a distance of $d \approx \qty{2}{\kilo\pc}$ away and rotates at $\nu \approx \qty{30}{\hertz}$. Thus,
\begin{equation}
    h_0 \approx \num{2e-27} \left( \frac{\epsilon}{\num{e-6}} \right) \left( \frac{I_3}{\qty{e45}{\g\cm\squared}} \right).
\end{equation}
The sceptic may be tempted to call it a day there; clearly such systems are well beyond current (and expected future) sensitivities of gravitational-wave instruments---surely this is a deeply misguided exercise? However, there is cause for some (cautious) optimism. If the source is observed for a sufficiently long period of time, the strain data measured at the detector can be folded over itself in order to improve the signal-to-noise ratio. The simplest version of this is matched filtering, where the effective strain amplitude $h_\text{c}$ increases as the square-root of the number of detected cycles. Suppose we could observe the Crab for a year. Then we have
\begin{equation}
    h_\text{c} \approx \num{e-22} \left( \frac{\epsilon}{\num{e-6}} \right) \left( \frac{I_3}{\qty{e45}{\g\cm\squared}} \right) \left( \frac{T_\text{obs}}{\qty{1}{\yr}} \right)^{1/2},
\end{equation}
where $T_\text{obs}$ is the length of the observing run. This is a substantial improvement.

But, so far, we have made no mention of how large these mountains might be. An obvious quantitative question we can ask is how large can we reasonably expect $\epsilon$ to be? This limit is set by the crust of the neutron star. This is obviously a crucial, but involved question, which we will discuss in Sec.~\ref{sec:Crust}. But for now, we will consider whether there are other types of observations that may provide some hope for this endeavour.

\subsection{Evidence from electromagnetic observations}
\label{sec:ElectromagneticEvidence}

\begin{figure}[h]
    \includegraphics[width=0.55\textwidth]{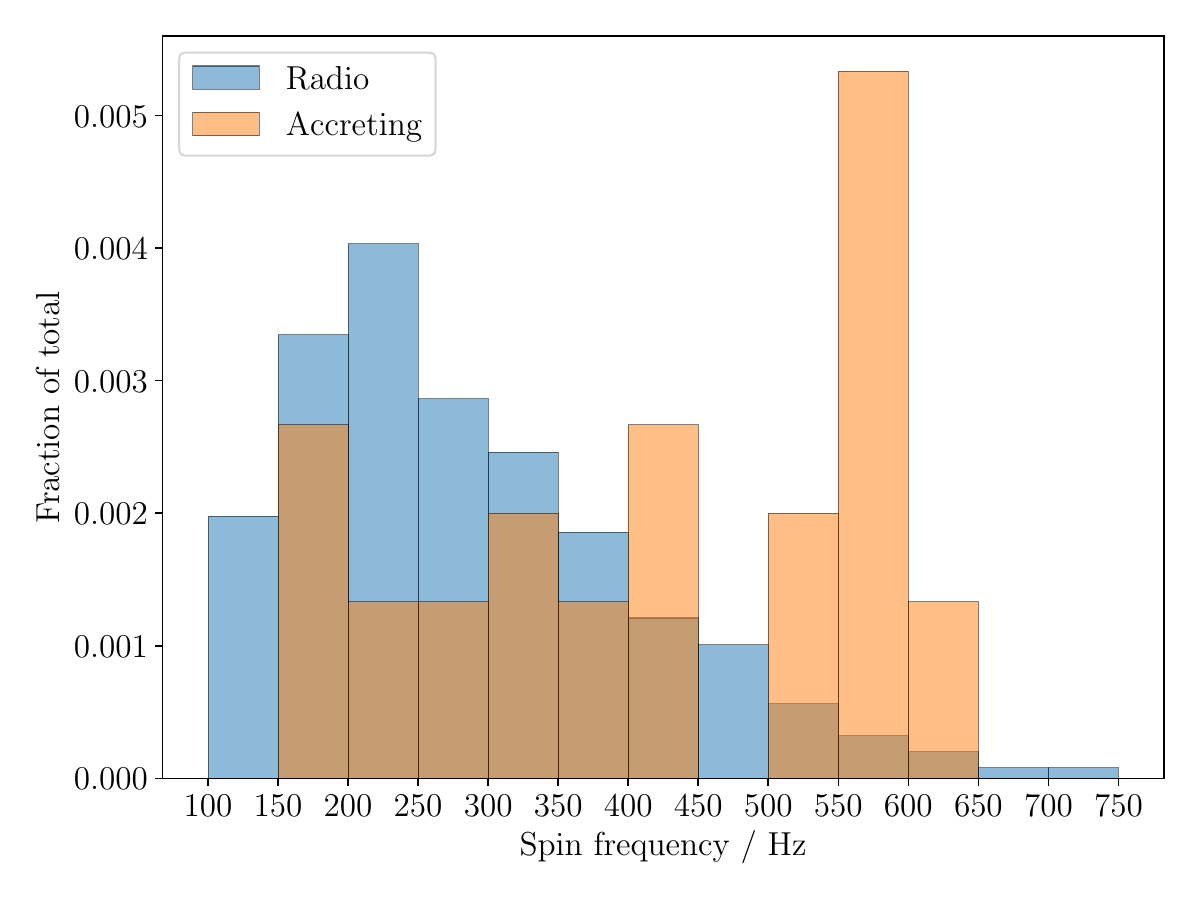}
    \caption{\label{fig:Distribution}%
    The distributions of spin frequencies above \qty{100}{\hertz} for known radio and accreting neutron stars. No system spins faster than PSR J1748--2446ad at \qty{716}{\hertz} and the accreting systems pile up around $\sim \qty{600}{\hertz}$. This provides evidence for a spin-down torque in the population of rapidly rotating neutron stars. The radio data are obtained from the ATNF catalogue%
        \footnote{The ATNF Pulsar Database website: \url{https://www.atnf.csiro.au/people/pulsar/psrcat/}.}
        \cite{2005AJ....129.1993M}.
    }
\end{figure}

We are used to observing the electromagnetic emission from rotating neutron stars. Indeed, the first evidence for the existence of neutron stars came from a radio observation \cite{1968Natur.217..709H}. To date, we have seen approximately 3000 pulsars. By accurately timing the signals, we are able to determine their spins to high precision. According to Eq.~\eqref{eq:StrainAmplitude}, the strain depends sensitively on the rotation of the star. In principle, a (non-axisymmetric) star rotating at \qty{300}{\hertz} will radiate waves with an amplitude $(300/30)^2 = 100$ times stronger than the Crab pulsar, all else being equal. Therefore, the more rapidly rotating systems are among the most promising for gravitational-wave detection. The fastest-spinning neutron stars are either found as radio pulsars or in low-mass X-ray binaries. The two observed distributions of spins $\geq \qty{100}{\hertz}$ are shown in Fig.~\ref{fig:Distribution}.

In particular, accreting neutron stars have long been considered potential gravitational-wave emitters \cite{1978MNRAS.184..501P,1984ApJ...278..345W,1998ApJ...501L..89B,1999ApJ...516..307A}. (We will discuss this scenario further in Sec.~\ref{sec:Accretion}.) The reason why is twofold. Firstly, accretion is expected to play an important role in producing fast rotators, as the accreted gas carries angular momentum to the neutron star \cite{1982Natur.300..728A,1982CSci...51.1096R}. The second is related to the star's quadrupole moment. As the matter falls from the accretion disc and gets close to the neutron star, the magnetic-field lines will direct the gas towards the poles. Assuming the magnetic poles are misaligned with the rotation axis (which is supported by observations of pulsars), the star will develop a natural mass asymmetry. This picture is somewhat simplistic, but it provides the key motivation.

It is believed that the rapidly rotating radio pulsars are spun up through the accretion scenario \cite[for a review on observations, see][]{2021ASSL..461..143P}. It has been suggested that accretion should be sufficient to spin up neutron stars to the centrifugal break-up frequency \cite{1994ApJ...424..823C}, which should be $\sim \qty{1}{\kilo\hertz}$ for most equations of state \cite{2007PhR...442..109L}. However, the fastest observed pulsar PSR J1748--2446ad rotates at \qty{716}{\hertz} \cite{2006Sci...311.1901H}, far below the mass-shedding limit. Additionally, there have been statistical studies that show the population as a whole possesses a cut-off around \qty{730}{\hertz} \cite[][\textit{cf.}, Fig.~\ref{fig:Distribution}]{2003Natur.424...42C,2005ASPC..328..279C,2010ApJ...722..909P,2017ApJ...850..106P}. This issue is not new and it has been argued that the apparent speed limit indicates the presence of a spin-down mechanism \cite{2018A&A...620A..69H}.

The star will be spun down by any mechanism that takes away angular momentum. There are two candidates to explain the observations. One is the interaction between the magnetic-field lines and the accretion disc \cite{1977ApJ...217..578G,1978ApJ...223L..83G,1979ApJ...232..259G,1979ApJ...234..296G,1997ApJ...490L..87W,2005MNRAS.361.1153A}. The second is the emission of gravitational waves \cite{1998ApJ...501L..89B,1999ApJ...516..307A}. However, it has been suggested that incorporating only the magnetic field-accretion disc coupling is not sufficient to explain the observations and gravitational radiation is indeed necessary \cite{2019MNRAS.488...99G}.

Along this direction, there is support for a \textit{minimum} quadrupole in the population of fast-spinning neutron stars. The recent study of \citet{2018ApJ...863L..40W} consider 128 millisecond pulsars with measurements of the stars' spin and spin-down rate. By considering a simple model for the spin behaviour of the neutron stars, incorporating electromagnetic dipole and gravitational-wave quadrupole radiation, they find strong statistical support for the population possessing a non-zero minimum ellipticity of $\epsilon \approx \num{e-9}$.

While gravitational waves are a natural and attractive explanation for features in the observed distribution (Fig.~\ref{fig:Distribution}), it is by no means a settled issue. The disc-magnetosphere interaction still remains a viable candidate for specific systems like XTE J1814--338 and SAX J1808.4--3658 \cite{2011ApJ...738L..14H} and some studies argue against the necessity of gravitational waves to describe the fast-spinning population as a whole \cite{2012ApJ...746....9P,2017MNRAS.470.3316D}. \cite[See][for a detailed discussion of the two perspectives.]{2017ApJ...850..106P} Recently, \citet{2021MNRAS.505L.112E} conjecture that the maximum spin frequency is explained by a correlation between the accretion rate and frozen magnetic field. Let us move on to consider more precisely how gravitational radiation impacts the dynamics.

From Eq.~\eqref{eq:Reaction}, a star with a quadrupole moment will radiate energy according to
\begin{equation}
    \frac{dE}{dt} = - \frac{32 G}{5 c^5} \epsilon^2 I_3^2 \Omega^6.
    \label{eq:RadiateEnergy}
\end{equation}
We assume that the moment of inertia $I_3$ is fixed, so the energy must come from the rotation, exerting a torque on the star. Thus, the star spins down according to
\begin{equation}
    \frac{d\Omega}{dt} = - \frac{32 G}{5 c^5} \epsilon^2 I_3 \Omega^5.
    \label{eq:SpinDown}
\end{equation}
This result shows that gravitational waves will cause the star to spin down proportional to $\Omega^5$. Therefore, the faster the star spins, the greater its spin-down from gravitational waves. This behaviour would support a speed limit provided the star hosts a non-trivial deformation $\epsilon$.

Alas, we do not expect realistic neutron stars to be so simple. In reality, a neutron star will have a variety of torques acting on it during its lifetime, which will influence its spin evolution. In Box~\ref{box:BrakingIndex}, we introduce the braking index $n$, which is often used to categorise pulsars. The Crab pulsar has been measured to have $n \approx 2.5$, which supports its spin being dominated by electromagnetic waves, as opposed to gravitational radiation \cite{2015MNRAS.446..857L}.

\begin{tbox}[label=box:BrakingIndex]{The braking index}
When pulsars are observed using radio or X-ray instruments, typically what is recorded are discrete measurements of the spin frequency $\nu$ as a function of time. Modern telescopes have extremely high resolution, so provided a clean enough signal the first $\dot{\nu}$ and second time derivatives $\ddot{\nu}$ can be inferred from this information.

There are a variety of different mechanisms that can influence the spin of a neutron star. The vast majority of pulsars are observed to be spinning down. It is therefore convenient to define the braking index $n$ by
\begin{equation*}
    \dot{\nu} \propto \nu^n \quad \implies \quad n = \frac{\nu \ddot{\nu}}{\dot{\nu}^2},
\end{equation*}
which can be calculated from the observational data. The most common mechanisms for spin-down torques in the literature are the following:
\begin{itemize}
    \item $n = 1$---relativistic particle wind \cite{1969ApJ...158..727M,1969Natur.223..277M},
    \item $n = 3$---electromagnetic dipole radiation \cite{1968Natur.219..145P},
    \item $n = 5$---gravitational mass-quadrupole radiation (a mountain),
    \item $n = 7$---gravitational current-quadrupole radiation \cite[\textit{e.g.}, an \textit{r}-mode;][]{1980RvMP...52..299T}.
\end{itemize}
Thus, accurately tracking the rotation rate can give an indication of the dominant torques on the star.

In practice, it can be challenging to get a clean signal. There are 12 systems for which there are reliable estimates of the associated braking indices \cite{2007Ap&SS.308..317L,2011MNRAS.411.1917W,2011ApJ...741L..13E,2011ApJ...742...31L,2012MNRAS.424.2213R,2015MNRAS.446..857L,2015ApJ...812...95F,2016ApJ...819L..16A,2016ApJ...832L..15C,2017MNRAS.466..147E}. They all cluster around $n = 3$ and below. However, there has been some evidence in support of gravitational-wave emission.

\citet{2020MNRAS.499.5986S} suggested that GRB 061121 may have a magnetar remnant powering the emission with a braking index of $n = 4.85^{+0.11}_{-0.15}$, which would imply the presence of a mass quadrupole. There has also been some discussion surrounding the inter-glitch evolution of PSR J0537--6910, which displays behaviour consistent with $n = 7$ \cite{2018ApJ...864..137A,2020MNRAS.498.4605H}.

It should be noted that the braking index is a measure of the dominant spin-down torque on the star. It is entirely consistent that a star may be predominantly spinning down due to the emission of electromagnetic radiation and still emit gravitational waves. Indeed, this was shown by \citet{2000A&A...354..163P} who derived non-trivial, upper limits on the ellipticity of four pulsars with $n < 3$.
\end{tbox}

But what about other pulsars? The accreting pulsar J1023+0038 is a particularly interesting source. It spins at \qty{592}{\hertz} and has been observed to transition between a radio state and an accretion-powered X-ray state \cite{2017PhRvL.119p1103H}. Timing the two phases shows that the neutron star spins down 27\% more rapidly during the accretion phase. This is surprising, since accretion is expected to typically spin the star up. However, this behaviour is consistent with the emergence of a mountain during the X-ray phase, resulting in a greater spin-down. This would require an ellipticity of $\epsilon \approx \num{5.7e-10}$ to explain the additional spin-down through gravitational-wave emission. Using a different approach and exploring a range of parameters, \citet{2020MNRAS.498..728B} revisited this source and found similar ellipticities. Additionally, \citet{2020PhRvD.102d3020C} provide estimates for 13 systems in the range of $\num{9e-10} \lesssim \epsilon \lesssim \num{2.34e-8}$.

\section{Crustal strains}
\label{sec:Crust}

We have seen how a non-axisymmetric, spinning star will radiate gravitational waves. The gravitational-wave strain that is measured by, \textit{e.g.}, an interferometer on Earth depends on the distance to the source $d$, how fast it rotates $\Omega$ and how strongly it is deformed $\epsilon$. We observe pulsars all across the galaxy with frequencies as high as $\nu \sim \qty{700}{\hertz}$.

The obvious place to start in considering deformations of neutron stars is to assume that there is strain in its solid crust. In this section, we will discuss the role of the crust in supporting asymmetries. We will see that the crust sets the limit on how large the mountain can be, but this comes with substantial uncertainties regarding how the mountain was formed in the first place.

\subsection{An estimate from energetics}
\label{sec:Energetics}

Shortly after the discovery of glitches, a crust-quake model was proposed \cite{1971AnPhy..66..816B,1972PEPI....6..103P}. Ultimately, this model went out of favour when observations required a larger energy budget than was physically feasible \cite{2012puas.book.....L}. Although this framework was unable to describe the glitch phenomenon, it proves to be quite useful for discussing likely asymmetries in neutron stars.

We return to the star of Fig.~\ref{fig:DeformedStar} with an ellipticity $\epsilon$ and assume it to be non-rotating for simplicity. Suppose that its crust would be unstrained if the star had a shape given by an ellipticity of $\epsilon_0$. The star then has energy
\begin{equation}
    E = E_0 + A \epsilon^2 + B (\epsilon - \epsilon_0)^2.
\end{equation}
The first term $E_0$ is the energy the star would have were it spherical and without a crust. The second contribution $A \epsilon^2$ represents the increase in gravitational potential energy due to its non-spherical shape $\epsilon$. The final term $B (\epsilon - \epsilon_0)^2$ is the energy stored up in the strain of the elastic crust. The quantities $A$ and $B$ are related to the star's binding energy and the shear modulus, respectively, and we assume them to be constant.

The shape $\epsilon$ of the star will minimise the energy, such that
\begin{equation}
    \epsilon = \frac{B}{A + B} \epsilon_0.
\end{equation}
If the star had no crust at all, then $B = 0$ and the ellipticity would vanish. A non-rotating, fluid star in the absence of other forces must be spherical. The opposite extreme is $A / B = 0$, corresponding to a rigid body. Then, we simply have $\epsilon = \epsilon_0$, where the elastic component holds the shape fixed.

Detailed calculations have found that $B / (A + B) \approx \num{2e-6}$ for quadrupole deformations \cite{2003ApJ...588..975C}. Stresses in the crust will only slightly deform the body because the Coulomb forces are much weaker than the strong gravity of the neutron star. Therefore, given a relaxed shape $\epsilon_0$, the neutron star will settle into an ellipticity of $\epsilon \approx \num{2e-6} \epsilon_0$, held in tension by the gravitational and elastic forces.

We can use this simple estimate to shed some light on how large the deformation can be. This is determined by the crust. Above a critical strain $\bar{\sigma}_\text{max}$, known as the \textit{breaking strain}, the crust yields and the strain is released. Therefore, we set $|\epsilon_\text{max} - \epsilon_0| = \bar{\sigma}_\text{max}$ to find
\begin{equation}
    \epsilon_\text{max} \approx \num{2e-7} \left( \frac{\bar{\sigma}_\text{max}}{0.1} \right).
    \label{eq:EllipticityEnergetics}
\end{equation}
There is some degree of uncertainty around how large the crustal breaking strain may be. Molecular-dynamic simulations for high-pressure Coulomb crystals suggest that it is remarkably strong, $\bar{\sigma}_\text{max} \approx 0.1$ \cite{2009PhRvL.102s1102H}. However, this is under the assumption that the crust is a perfect crystal. Indeed, \citet{2018MNRAS.480.5511B} argue that neutron-star crusts will have lower strains, $\bar{\sigma}_\text{max} \approx 0.04$. For more amorphous materials, like many terrestrial solids, the breaking strain will lie in the range $\num{e-4} \leq \bar{\sigma}_\text{max} \leq \num{e-2}$. For a neutron star of characteristic radius $R \sim \qty{10}{\km}$, these deformations are of the size $\sim \epsilon_\text{max} R \sim \qty{0.1}{\cm}$. So they are unlikely to rival the peaks we find on Earth and perhaps it is rather grandiose to call them ``mountains''.

\subsection{Elastic deformations}
\label{sec:Elastic}

If we want to be more precise---which we must in order to say anything quantitative about the physics---we need to consider the stellar interior. As we will see, this is not a totally straightforward venture; it involves evolutionary aspects of neutron stars that are currently poorly understood. In particular, we need to know why the star is deformed in the first place and what the relaxed shape $\epsilon_0$ is likely to be for a typical neutron star. As we have discussed, this will ultimately be a challenging task, but necessary to make progress.

We consider a neutron star with mass density $\rho$, pressure $p$ and gravitational potential $\Phi$. The star is governed by the following system of equations \cite[for a modern textbook, see][]{2017mcp..book.....T}: mass conservation (the continuity equation)
\begin{subequations}\label{eqs:Full}
\begin{equation}
    \partial_t \rho + \nabla_i (\rho v^i) = 0,
\end{equation}
momentum conservation (the Euler equation with shear stresses)
\begin{equation}
    \rho \frac{dv_i}{dt} = - \nabla^j (p \, g_{i j} + t_{i j}) - \rho \nabla_i \Phi
    \label{eq:ForceFull}
\end{equation}
and the field equation (Poisson's equation)
\begin{equation}
    \nabla^2 \Phi = 4 \pi G \rho,
\end{equation}
\end{subequations}
where $v^i$ is the velocity field, $t_{i j}$ is the shear-stress tensor of the crust, $g_{i j}$ is the flat-space, Euclidean metric, $\nabla_i$ is a covariant derivative and $\nabla^2 = \nabla_i \nabla^i$ is the Laplacian. The crust enters the system in Eq.~\eqref{eq:ForceFull} through the star's stress tensor $(p \, g_{i j} + t_{i j})$. The presence of a crust enables the star to support anisotropic stresses and to deviate from axisymmetry. In practice, Eqs.~\eqref{eqs:Full} are difficult to solve, but there are some convenient simplifications we can make.

First, we assume that the star does not deviate far from sphericity. That is, we can construct a spherically symmetric, unstrained background from the usual equations of stellar structure
\begin{subequations}\label{eqs:Background}
\begin{gather}
    \nabla_i p = - \rho \nabla_i \Phi, \\
    \nabla^2 \Phi = 4 \pi G \rho
\end{gather}
\end{subequations}
and the asymmetric contributions are calculated as perturbations on top of this background. We denote an Eulerian perturbation of a quantity by $\delta$. Second, we will assume that the perturbations are static. By considering the difference of Eqs.~\eqref{eqs:Full} and \eqref{eqs:Background} and linearising, we obtain the following perturbation equations \cite{1967MNRAS.136..293L,1978ApJ...221..937F}:
\begin{subequations}\label{eqs:Perturbations}
\begin{gather}
    \delta \rho = - \nabla_i (\rho \xi^i), \\
    0 = - \nabla^j (\delta p \, g_{i j} + t_{i j}) - \delta \rho \nabla_i \Phi - \rho \nabla_i \delta \Phi, \label{eq:Force}\\
    \nabla^2 \delta \Phi = 4 \pi G \, \delta \rho,
\end{gather}
\end{subequations}
where $\xi^i$ is the Lagrangian displacement vector that describes the position of fluid elements relative to their position in the spherical configuration. Third and finally, we will assume that the perturbations are entirely in the $(l, m) = (2, 2)$ spherical harmonic, such that, \textit{e.g.}, $\delta \rho(r, \theta, \phi) = \delta \rho_{2 2}(r) \operatorname{Re}[Y_2^2(\theta, \phi)]$. This is obviously quite idealised. A realistic evolutionary history will likely result in the star being deformed in other harmonics, but the $(l, m) = (2, 2)$ harmonic couples the most strongly with gravitational waves.%
\footnote{As we saw in Sec.~\ref{sec:GravitationalWaves}, the leading-order contribution to the strain comes from the quadrupole moment tensor [see Eq.~\eqref{eq:StrainQuadrupole}]. In general, if there is also an $(l, m) = (2, -2)$ perturbation, this will also contribute to the gravitational-wave strain. However, since we are interested in real, physical perturbations of the star, this has identical angular dependence to the $(l, m) = (2, 2)$ harmonic.}
We note that other harmonics will contribute to the strain of the crust, without increasing the quadrupole moment. This simplification is suitable for considering upper limits on the size of the deformations.

There exists a simple relationship between the ellipticity $\epsilon$ and the $(l, m) = (2, 2)$ moment (the quadrupole moment) 
\begin{equation}
    Q_{2 2} = \int \delta \rho_{2 2}(r) r^4 \, dr
\end{equation}
(see Box~\ref{box:Multipole}). Given that the perturbations describe the deviation from sphericity, we have [\textit{cf.} Eq.~\eqref{eq:Ellipticity}]
\begin{equation}
    \epsilon = \sqrt{\frac{8 \pi}{15}} \frac{Q_{2 2}}{I_3},
    \label{eq:FiducialEllipticity}
\end{equation}
where $I_3$ is sourced by the spherical background. It is common to assume the fiducial value $I_3 = \qty{e45}{\g\cm\squared}$. Under this assumption, $\epsilon$ is known as the \textit{fiducial ellipticity}, which is often reported in observational papers. The star's true principal moment of inertial can be different from this fiducial value by a factor of a few. There are no particular difficulties in calculating $I_3$; however, this convention exists and it is more than sufficient to make order-of-magnitude estimates.

\begin{figure}[h]
    \includegraphics[width=0.5\textwidth]{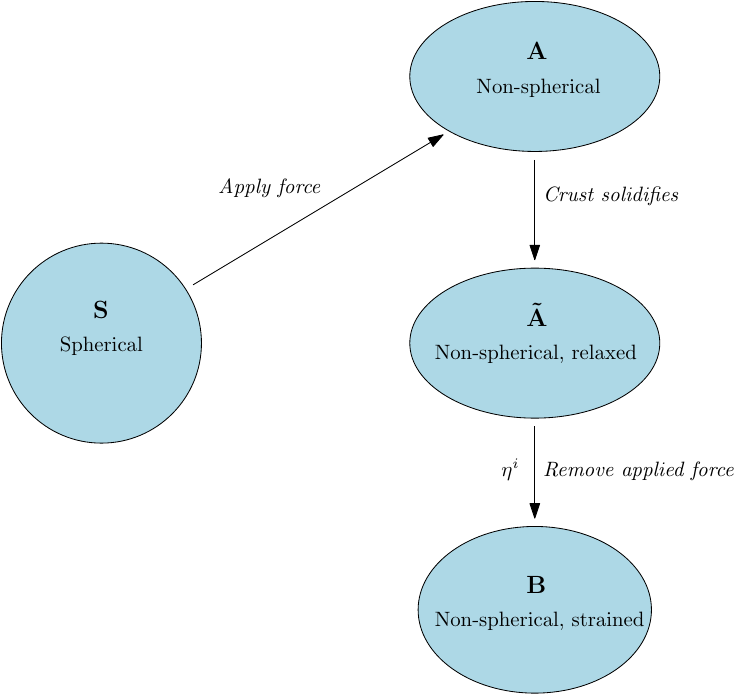}
    \caption{\label{fig:Stars}%
    A schematic illustration of the evolutionary history of a deformed neutron star \cite[reproduced from][]{2021MNRAS.500.5570G}. The spherical star S is simply a reference shape used to define the quadrupole moment and construct perturbations with respect to. It is not a shape the star necessarily ever had. Star {\~A} is the shape the crust is relaxed in. The abstract force maps star A (and star {\~A}) to star S. The force represents the formation history of the history that leads to the non-spherical, unstrained shape; it is not a physical, external force acting on the star. The final configuration, star B, supports the asymmetric deformations self-consistently through the elastic strains described by the displacement vector $\eta^i$.
    }
\end{figure}

We need to be careful regarding the crustal strains that sit in Eq.~\eqref{eq:Force}. The shear-stress tensor is given by
\begin{equation}
    t_{i j} = - 2 \hat{\mu} \sigma_{i j},
    \label{eq:Stress}
\end{equation}
where $\hat{\mu}$ is the shear modulus of the crust and
\begin{equation}
    \sigma_{i j} = \frac{1}{2} (\nabla_i \eta_j + \nabla_j \eta_i) - \frac{1}{3} \nabla_k \eta^k g_{i j}
    \label{eq:Strain}
\end{equation}
is the symmetric, trace-free strain that arises from a displacement field $\eta^i$. We expect a typical neutron star to have a complicated evolutionary history such that its unstrained shape---the shape the crust will be relaxed in---will not be spherical (see Fig.~\ref{fig:Stars}). (The star has an ellipticity $\epsilon$, whereas the relaxed configuration has an ellipticity $\epsilon_0$, connecting with the discussion of Sec.~\ref{sec:Energetics}.) Since, in general, we cannot assume that the relaxed shape of the crust is that of a perfect sphere, we assert $\eta^i \neq \xi^i$.

\textit{A priori}, we do not know what the relaxed configuration of the star is. Nevertheless, we can make progress by considering how large a mountain can be before the neutron-star crust breaks. We will see that there are subtleties in this question. The earliest study in this direction was by \citet{2000MNRAS.319..902U}, assuming the Cowling approximation---neglecting perturbations of the gravitational potential $\delta \Phi = 0$. Under this approximation, there is no support for pressure perturbations in the fluid and only the crust contributes to the quadrupole moment. The maximum mountain will occur when the crust is close to breaking. To define the elastic limit, it is common to use the von Mises criterion: the solid will yield when the von Mises strain $\bar{\sigma}$, given by
\begin{equation}
    \bar{\sigma}^2 = \frac{1}{2} \sigma_{i j} \sigma^{i j},
    \label{eq:vonMises}
\end{equation}
exceeds the breaking strain $\bar{\sigma} \geq \bar{\sigma}_\text{max}$. \citet{2000MNRAS.319..902U} imposed that the crust was maximally strained at every point, subject to the criterion~\eqref{eq:vonMises}. With this shape, they obtained a quadrupole moment of
\begin{subequations}\label{eqs:UCB}
\begin{equation}
     Q_{2 2}^\text{max} \approx \num{1.2e39} \left( \frac{\bar{\sigma}_\text{max}}{0.1} \right) \left( \frac{1.4 M_\odot}{M} \right)^{1.2} \left( \frac{R}{\qty{10}{\km}} \right)^{6.26} \, \unit{\g\cm\squared}.
\end{equation}
In terms of the fiducial ellipticity~\eqref{eq:FiducialEllipticity},
\begin{equation}
    \epsilon_\text{max} \approx \num{1.6e-6}  \left( \frac{\bar{\sigma}_\text{max}}{0.1} \right) \left( \frac{1.4 M_\odot}{M} \right)^{1.2} \left( \frac{R}{\qty{10}{\km}} \right)^{6.26}.
\end{equation}
\end{subequations}

The maximum mountain of \citet{2000MNRAS.319..902U} differs from the simple energetics argument~\eqref{eq:EllipticityEnergetics} by a factor of a few. However, there are a number of simplifications embedded in the result---in particular, the Cowling approximation and Newtonian gravity. These two assumptions were lifted by \citet{2013PhRvD..88d4004J}. For an $M = 1.4 M_\odot$ star described by the SLy equation of state \cite{2001A&A...380..151D}, they obtained
\begin{equation}
    Q_{2 2}^\text{max} \approx \num{2e39}  \left( \frac{\bar{\sigma}_\text{max}}{0.1} \right) \, \unit{\g\cm\squared}, \qquad \epsilon_\text{max} \approx \num{3e-6} \left( \frac{\bar{\sigma}_\text{max}}{0.1} \right).
\end{equation}

It was noted by \citet{2006MNRAS.373.1423H} \cite[and later discussed in more detail by][]{2021MNRAS.500.5570G} that, by demanding the crust to be at breaking strain throughout, the neutron star is forced into a shape that violates physical boundary conditions. The (linearised) traction vector $T^i$, given by [see Eq.~\eqref{eq:Force}] 
\begin{equation}
    T^i = (\delta p \, g^{i j} + t^{i j}) \nabla_j r,
    \label{eq:Traction}
\end{equation}
must be continuous throughout the star. At an interface between a fluid region (the neutron-star core or ocean) and the solid crust, there is expected to be a first-order phase transition where the crust sharply obtains a non-zero shear modulus $\hat{\mu}$. (In the fluid, this is zero, since it cannot support shear stresses.) In order for the traction~\eqref{eq:Traction} to be continuous, components of the strain tensor $\sigma_{i j}$ must go to zero at an interface. Since the crust is maximally strained at every point, the strain components have instead finite values. Thus, continuity of the traction cannot be satisfied and the maximally strained approach is unphysical.

One could imagine allowing the star to be at breaking strain at every point except at the interfaces, such that the traction is continuous. This \textit{ad hoc} situation is difficult to implement and the result will likely be very similar to that of \citet{2000MNRAS.319..902U}. However, these issues raise the important question of whether we expect the crust to get so close to breaking strain throughout?

\citet{2006MNRAS.373.1423H} considered the situation with an $(l, m) = (2, 2)$ force at the surface that supported the deformation. They calculated Newtonian stellar models, ensuring continuity of the traction, and found that---under such a surface force---the crust yielded at specific points. The corresponding results were (surprisingly) an order of magnitude more optimistic than that of Eqs.~\eqref{eqs:UCB}. But, in their calculation, they implicitly assumed the relaxed shape to be spherical, where $\eta^i = \xi^i$ in the strain~\eqref{eq:Strain}.

In order to capture the deformed, relaxed configuration, it turns out to be convenient to introduce an abstract force density into the problem \cite{2021MNRAS.500.5570G}. The situation is illustrated schematically in Fig.~\ref{fig:Stars}. The force density contains information about the non-spherical shape the crust will be relaxed in and is therefore related to the star's evolutionary history. But, it plays no role in supporting the mountain; the deformed shape of the star is self-consistently supported by crustal strains and a solution to Eqs.~\eqref{eqs:Perturbations}.

\begin{figure}[h]
    \includegraphics[width=0.6\textwidth]{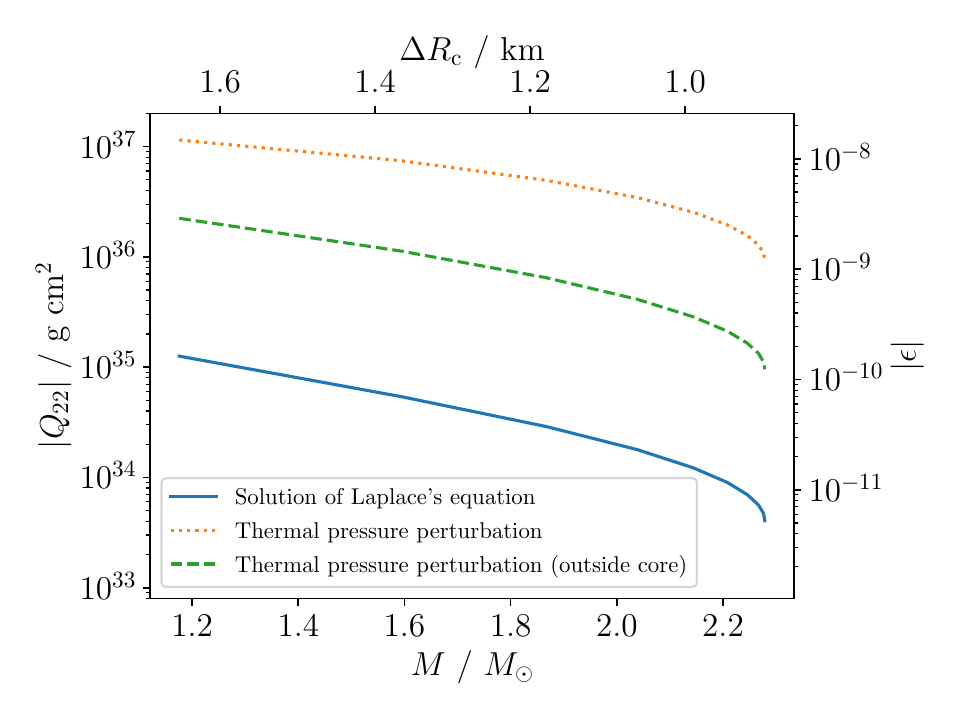}
    \caption{\label{fig:Results}%
    The maximum quadrupole deformation, given in terms of the quadrupole moment $Q_{2 2}$ and fiducial ellipticity $\epsilon$, due to different perturbing forces as functions of the stellar mass $M$ and crustal thickness $\Delta R_\text{c}$ \cite[reproduced from][]{2021MNRAS.507..116G}. The perturbing forces are simple mathematical models for illustrative purposes with no explicit connection to the neutron-star physics. The results illustrate that the size of the mountain strongly depends on the neutron star's formation history.
    }
\end{figure}

\citet{2021MNRAS.500.5570G} suggested a scheme for calculating the structure of the strained star with knowledge of the unstrained shape through the abstract force. The inclusion of the abstract force is important to ensure the boundary conditions are satisfied, such that the shape of the star is a physically acceptable solution. \citet{2021MNRAS.507..116G} used this scheme to calculate the mountains of fully relativistic neutron-star models subject to some mathematically simple forms of the abstract force. The results, displayed in Fig.~\ref{fig:Results}, show the dependence of the mountain on the history of the neutron star. This has been further supported by the recent (Newtonian) calculations of \citet{2022MNRAS.517.5610M}.

Moving forward, we will inevitably require detailed, evolutionary simulations of neutron stars in order to understand how such deformations can arise. Even for a simple, isolated neutron star, there is a lot of physics to keep track of. To simulate such a system, one would need to model the crust, rapid rotation, electromagnetic dipole radiation and glitches. However, aspects of the physics are highly uncertain. Indeed, it is generally assumed that, once the elastic crust reaches the yield point, all the strain is released. It is not clear how appropriate this assumption is. Additionally, both the magnetic-field configuration of a typical neutron star and reliable glitch models are open questions. Along this vein, \citet{2022MNRAS.514.1628K} developed a phenomenological model where asymmetries are developed as the star spins down due to dipole radiation and the crust repeatedly fractures. They find that the star needs to be born rapidly spinning with $\gtrsim \qty{750}{\hertz}$ for this process of crustal fracturing to occur and obtain very modest final ellipticities in the range of $\num{5e-14} \lesssim \epsilon \lesssim \num{e-12}$.

It will also be interesting to consider the effects of plasticity. Calculations generally assume the crust to be an elastic solid, possessing the linear relationship of Eq.~\eqref{eq:Stress} between the strain and the induced stresses. Terrestrial materials tend to exhibit some form of plastic behaviour and neutron-star crusts are likely to be similar \cite{1970PhRvL..24.1191S,2003ApJ...595..342J,2010MNRAS.407L..54C}. If this is the case, then the crust will retain some of the strain beyond the yield point and the star may be able to develop larger quadrupole moments.

Before we conclude our discussion on crustal deformations, it is worth noting that the maximum ellipticities presented so far assume the neutron stars to be composed of ordinary baryonic matter. Given the extreme pressures in neutron-star cores, it is conceivable that more exotic phases of matter exist \cite[for a review, see][]{2018RPPh...81e6902B}. \citet{2013PhRvD..88d4004J} considered additional phases and obtained maximum quadrupoles of $\sim \qty{e44}{\g\cm\squared}$ for $1.4 M_\odot$ entirely crystalline colour-superconducting quark stars and $\sim \qty{e41}{\g\cm\squared}$ for $2 M_\odot$ hadron-quark hybrid stars with solid cores. These more exotic stars can sustain significantly larger quadrupole deformations before their crust yields.

\section{The magnetic field}
\label{sec:Magnetic}

So far, we have paid little attention to what physical processes may produce the mountain. A natural place to start is to consider the role of the magnetic field, which is particularly strong in neutron stars. It was first shown by \citet{1953ApJ...118..116C} that magnetic fields on stars source quadrupolar deformations. Given that we observe neutron stars that host magnetic fields misaligned with their spin axis, these are natural sources of gravitational radiation. In addition, the remarkable observations of the NICER mission indicate that neutron stars may have rich, complex magnetic-field structures beyond the standard dipole assumptions \cite{2019ApJ...887L..21R,2019ApJ...887L..23B,2019ApJ...887L..24M,2021ApJ...918L..27R,2021ApJ...918L..28M}. Here, we discuss the simple case of an isolated neutron star with a magnetic field that supports a mountain. We call these \textit{magnetic mountains}. We will delay discussing the role of the magnetic field in accretion until Sec.~\ref{sec:Confinement}.

Another simple energetics argument leads to \cite{2008MNRAS.385..531H}
\begin{equation}
    \epsilon \sim \frac{B^2 R^3}{G M^2 / R} \approx \num{e-12} \left( \frac{R}{\qty{10}{\km}} \right)^4 \left( \frac{1.4 M_\odot}{M} \right)^2 \left( \frac{B}{\qty{e12}{\gauss}} \right)^2,
    \label{eq:MagneticEllipticity}
\end{equation}
where $B$ represents the magnetic-field strength. Here, we have compared the energy stored in the magnetic field to the star's gravitational potential energy. Similar to as we saw for the crust in Sec.~\ref{sec:Energetics}, the neutron star's self-gravity is very strong compared to the magnetic field.

However, this estimate of the field strength is based on the \textit{external} magnetic field that is inferred from pulsar timing. The interior field, which will be more relevant for the shape of the star, is highly uncertain and is (quite predictably) a complicated problem. The main theoretical challenge that we face is our present inability to construct stable models. Both analytical \cite{1956ApJ...123..498P,1973MNRAS.161..365T} and numerical studies \cite{2006A&A...453..687B,2007A&A...469..275B} indicate that purely poloidal and purely toroidal magnetic fields are unstable on dynamical timescales. Whereas mixed fields, where the toroidal field threads the closed-field-line region of the poloidal component (a twisted-torus configuration), may be stable \cite{2006A&A...450.1077B,2008MNRAS.383.1551A,2009MNRAS.397..913C}, but this could be equation-of-state dependent \cite{2012MNRAS.424..482L,2015MNRAS.447.1213M}. Of course, much of our present understanding involves idealised assumptions, like ideal magnetohydrodynamics and barotropic equations of state. It could also be that a real neutron star is not in always in equilibrium---the magnetic field may be gradually evolving.

Despite this, there have been calculations of the kind of deformations that a magnetic field on a neutron star may sustain \cite{1996A&A...312..675B,2002PhRvD..66h4025C,2008MNRAS.385..531H,2008CQGra..25k4049H,2010MNRAS.406.2540C,2012PhRvL.109h1103G,2019MNRAS.490.2692K,2020MNRAS.499.2636B,2021A&A...654A.162S}. These more detailed calculations approximately agree with the order-of-magnitude estimate of Eq.~\eqref{eq:MagneticEllipticity}, which is not particularly optimistic from a gravitational-wave standpoint. Assuming the star has a superconducting phase and a purely toroidal field leads to the ellipticity scaling as \cite{2002PhRvD..66h4025C,2012PhRvL.109h1103G}
\begin{equation}
    \epsilon \sim \num{e-9} \left( \frac{B}{\qty{e12}{\gauss}} \right) \left( \frac{H_\text{c}}{\qty{e15}{\gauss}} \right),
\end{equation}
where $H_\text{c}$ is the critical field strength characterising superconductivity.

It should be further noted that, in principle, a mountain sustained by the magnetic field could be larger than what the crust allows. Indeed, a sufficiently large deformation would break the crust, but the magnetic field would hold the star in shape. A strong magnetic field, like that of a magnetar, will enable the star to develop a large ellipticity. But there is an important caveat to this. Suppose the magnetic field was $B \sim \qty{e15}{\gauss}$. This would lead to an ellipticity $\epsilon \sim \num{e-6}$, similar to what we saw for the maximum mountains sustained by crustal strains in Sec.~\ref{sec:Crust}. The downside is that a strong magnetic field would rapidly spin down the star due to electromagnetic radiation and reduce the gravitational-wave strain. It is conceivable that some fraction of neutron stars are born magnetars \cite{1998Natur.393..235K,2006csxs.book..547W} and, if their initial rotation rates are high enough, then newly born magnetars could be detectable sources of gravitational waves \cite{2001A&A...367..525P,2009MNRAS.398.1869D}. Clearly, this involves a number of assumptions and we do not know at present how likely this may be. The magnetars we observe and classify are consistently slow rotators.

This is not all the whole story for the magnetic field. It plays an important role in accreting neutron stars. This is the situation we go on to consider now.

\section{Accretion}
\label{sec:Accretion}

In our discussion of electromagnetic observations (Sec.~\ref{sec:ElectromagneticEvidence}), we alluded to the fact that accretion is expected to have a natural involvement in giving rise to non-axisymmetric deformations. In this context, the star is no longer isolated; it is joined by a companion star from which it is accreting gas. The gas will transfer from the gravitational influence of the companion to that of the neutron star, carrying some angular momentum, and form an accretion disc around it. The ionised gas in the disc will be channelled onto the surface following the magnetic-field lines. This process is inherently asymmetric and where the gas will land on the surface will depend on the configuration of the magnetic field. In the context of accretion, it is common to consider the balance between accretion torques, which generally spin the star up, and gravitational-wave torques, which will always spin the star down. This is discussed in brief in Box~\ref{box:Torque}.

\begin{tbox}[label=box:Torque]{Torque balance}
One can obtain an estimate for the quadrupole deformation of a neutron star in a low-mass X-ray binary by assuming that the angular momentum transferred via accretion from a companion is balanced by the radiation of gravitational waves \cite{1984ApJ...278..345W,1998ApJ...501L..89B}. One can approximate the spin-up torque due to accretion as \cite{1978ApJ...223L..83G}
\begin{equation*}
    \dot{J} \approx \dot{M} \sqrt{G M R},
\end{equation*}
where $\dot{M}$ is the rate of mass transfer to the neutron star. The energy radiated away as gravitational waves is given by Eq.~\eqref{eq:RadiateEnergy}, which will spin down the star. Balancing the two torques requires a quadrupolar deformation of
\begin{equation*}
    \epsilon \approx \num{1.3e-8} \left( \frac{\dot{M}}{\num{e-9} M_\odot \, \unit{\per\yr}} \right)^{1/2} \left( \frac{\qty{500}{\hertz}}{\nu} \right)^{5/2} \left( \frac{M}{1.4 M_\odot} \right)^{1/4} \left( \frac{R}{\qty{10}{\km}} \right)^{1/4} \left( \frac{\qty{e45}{\g\cm\squared}}{I_3} \right),
\end{equation*}
or equivalently $Q_{2 2} \approx \qty{9.8e36}{\g\cm\squared}$. Under this assumption, the most rapidly accreting systems will build the largest deformations. This is somewhat intuitive.

Now in reality, this is overly simplified. Accretion torques are more complicated and we have no particular reason to assume that the torques will balance. What this example does show is that relatively modest quadrupoles are required to influence the spin evolution of fast-spinning, accreting systems. However, this is particularly sensitive to the spin rate of the star. If the star is not fast spinning, then higher deformations are needed in order to produce significant gravitational-wave torques. In practice, the torque-balance estimate provides a useful number to benchmark quadrupole-forming mechanisms against.
\end{tbox}

\subsection{Thermal reactions}
\label{sec:Thermal}

The accretion scenario has received considerable attention when it comes to sourcing long-lived, quadrupole deformations on neutron stars \cite{2008MNRAS.389..839W,2015MNRAS.450.2393H}. \citet{1998ApJ...501L..89B} proposed that the accreted matter heats the crust giving rise to temperature-sensitive nuclear reactions involving electron captures. Hotter regions of the crust would have the reactions at lower pressures, so the density variations would occur at higher altitudes in these regions. This naturally gives rise to a quadrupole moment, which is commonly referred to as a \textit{thermal mountain}. There have since been developments upon this work, mostly in an effort to provide more detail to the modelling \cite{2000MNRAS.319..902U,2020MNRAS.493.3866S,2020MNRAS.494.2839O,2023MNRAS.522..226H}. Indeed, the problem is quite involved and requires an understanding of the composition of the neutron star.

A simple estimate for the quadrupole formed from nuclear reactions in the outer crust is given by \cite{2000MNRAS.319..902U}
\begin{equation}
    Q_{2 2} \approx \num{1.3e35} \left( \frac{R}{\qty{10}{\km}} \right)^4 \left( \frac{\delta T_\text{q}}{\qty{e5}{\kelvin}} \right) \left( \frac{E_\text{th}}{\qty{30}{\mega\electronvolt}} \right)^3 \, \unit{\g\cm\squared},
\end{equation}
where $\delta T_\text{q}$ is the quadrupole component of the temperature variation and $E_\text{th}$ is the reaction threshold energy. Higher threshold energies correspond to reactions deeper in the crust. The quadrupolar heating $\delta T_\text{q}$ will be a fraction of the total heating $\delta T$ due to the accreted mass $\Delta M$ \cite{2001MNRAS.325.1157U},
\begin{equation}
    \delta T \sim \num{2e5} \left( \frac{k_\text{B}}{C} \right) \left( \frac{\qty{e30}{\dyn\per\cm\squared}}{p_\text{d}} \right) \left( \frac{Q}{\qty{1}{\mega\electronvolt}} \right) \left( \frac{\Delta M}{\num{e-9} M_\odot} \right) \, \unit{\kelvin},
\end{equation}
where $C$ is the heat capacity per baryon, $p_\text{d}$ is the pressure in the star at which the energy is deposited and $Q$ is the (locally) released heat by the reactions per accreted baryon. The fraction of total heating that is quadrupolar is poorly understood, but estimated to be $\delta T_\text{q} / \delta T \lesssim 0.1$ \cite{2000MNRAS.319..902U}.

Particularly relevant for sustaining the mountain is how long the system is in outburst. Low-mass X-ray binaries are well known to exhibit periods of outburst, where the measured X-ray luminosity is high (implying rapid accretion rates), and periods of quiescence, where the luminosity is orders of magnitude lower. During low-accretion phases, the mountains are expected to decay on the crust's thermal timescale \cite{1998ApJ...504L..95B}
\begin{equation}
    \tau_\text{th} \approx 0.2 \left( \frac{1.4 M_\odot}{M} \right)^2 \left( \frac{R}{\qty{10}{\km}} \right)^4 \left( \frac{p_\text{d}}{\qty{e30}{\dyn\per\cm\squared}} \right)^{3/4} \, \unit{\yr}.
\end{equation}
If quiescence lasts longer than $\tau_\text{th}$, then the deformation will be washed away by gravity and a new mountain will be constructed during the next outburst. But a shorter recurrence time may, on the other hand, lead to an accumulation of material such that the mountain gets larger and larger. Although it is conceivable that the change in composition due to reactions in the initial outburst may be frozen and the mountain will get gradually larger.

\subsection{Magnetic confinement}
\label{sec:Confinement}

Another way an accreting neutron star is expected to develop an asymmetry is through the magnetic field confining the accreted matter \cite{1998ApJ...496..915B,2003AIPC..686...92P,2005ApJ...623.1044M,2007MNRAS.376..609P,2008MNRAS.386.1294V,2009MNRAS.395.1985V,2010MNRAS.402.1099W,2011MNRAS.417.2696P,2022MNRAS.516.5196F}. Accretion will interact with the magnetic field and compress it, such that there will be locally strong fields on the star's surface. This can lead to larger quadrupoles than those sourced by the background magnetic-field configuration, which we discussed previously in Sec.~\ref{sec:Magnetic}. We refer to these perturbations as \textit{magnetically confined mountains}.%
\footnote{At this point, the nomenclature can get somewhat clumsy. It is not uncommon in the literature to refer to these deformations as magnetic mountains, but in this review we will reserve this term for mountains solely sourced by the magnetic field.}

One of the differences with respect to thermal mountains is that the timescale on which the deformation relaxes is given by the Ohmic dissipation $\tau_\text{O} \geq \qty{e8}{\yr}$ \cite{2009MNRAS.395.1985V}. This is sufficiently long that a mountain will form over multiple outburst cycles. A simple approximation for the magnetic-field quenching leads to \cite{1989Natur.342..656S,2015MNRAS.450.2393H}
\begin{equation}
    Q_{2 2} \approx \num{e36} A \left( \frac{\Delta M}{\num{e-9} M_\odot} \right) \left( 1 + \frac{\Delta M}{M_\text{c}} \right)^{-1} \, \unit{\g\cm\squared},
\end{equation}
where $A \approx 1$ is a correction factor, which depends on the equation of state and the geometry of the accretion \cite{2005ApJ...623.1044M}, and $M_\text{c}$ is the critical accreted mass at which the mechanism supporting the mountain saturates, which also depends on the equation of state \cite{2011MNRAS.417.2696P}. Estimates for $M_\text{c}$ have been obtained by \citet{2004MNRAS.351..569P}, \citet{2005ApJ...623.1044M} and \citet{2011MNRAS.417.2696P}. There are numerical challenges to calculating for $\Delta M > M_\text{c}$. The size of the mountain depends strongly on the magnetic-field strength at the beginning of the accretion episode.

Early calculations for the mountains supported by the compressed magnetic field, such as \citet{2004MNRAS.351..569P} and \citet{2005ApJ...623.1044M}, assumed soft, isothermal equations of state for neutron-star matter. Under these assumptions, \citet{2005ApJ...623.1044M} obtained promising estimates for the magnitude of the deformations, up to $\epsilon \sim \num{e-4}$. Exploring the matter effects, \citet{2011MNRAS.417.2696P} found that the perturbations dropped to $\epsilon \approx \num{9e-7}-\num{e-8}$, depending on the precise equation of state.

Related to this mechanism is the possibility of fall-back accretion following a supernova or merger \cite{2014ApJ...794..170M,2015ApJ...798...25D,2021MNRAS.502.4680S}. In these scenarios, the accretion rates may be extremely high, so there is the added complication of the neutron star collapsing to form a black hole at some point. This situation, although interesting, will not produce sustained gravitational-wave emission.

\section{Precession}
\label{sec:Precession}

We have considered the most plausible mechanisms that will produce mountains on neutron stars. Before we go on to discuss the prospects for gravitational-wave detection, we will return to the simple model of a quadrupolar, spinning star from Sec.~\ref{sec:Rotating}.

\begin{figure}[h]
    \includegraphics[width=0.55\textwidth]{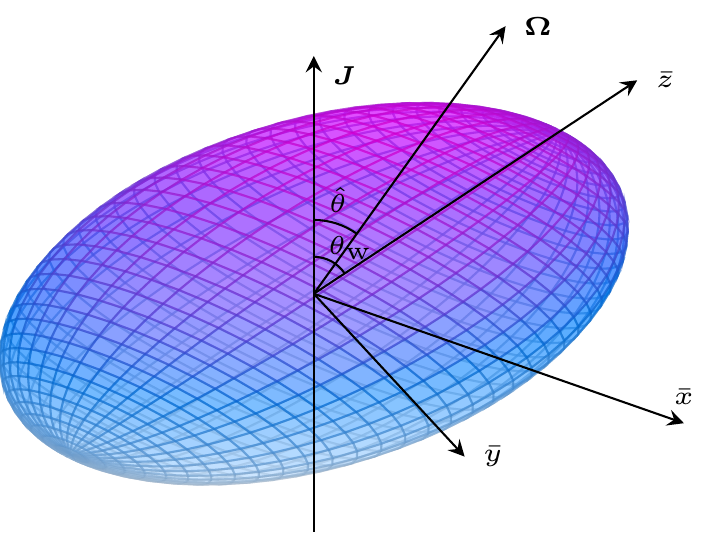}
    \caption{\label{fig:Precession}%
    An illustration of a deformed star with angular velocity $\bm{\Omega}$ precessing about its angular momentum $\bm{J}$. The star's body frame $(\bar{x}, \bar{y}, \bar{z})$ is shown along with two angles: $\theta_\text{w}$ and $\hat{\theta}$ are the angles between $\bm{J}$ and $\bm{\Omega}$ and $\bm{J}$ and the $\bar{z}$-axis, respectively. Both angles exist in the same plane throughout precession. This particular star is prolate in shape and symmetric with respect to the $\bar{z}$-axis.
    }
\end{figure}

Before, we assumed that the rotation of the star was fixed about an axis with respect to an inertial observer away from the star (as shown in Fig.~\ref{fig:DeformedStar}). However, the most general form of rotation is torque-free precession. The configuration of a freely precessing star is shown in Fig.~\ref{fig:Precession}. We will discuss the problem in detail below, but the central idea is the following: a body rotates about an axis, which in turn rotates about another axis. As we will see, contrary to the situation we considered prior, in free precession there are two components of the angular velocity (rather than one) and the associated gravitational waves radiate at an additional frequency.

The involvement of precession in gravitational radiation was first discussed by \citet{1978Natur.271..524Z}, \citet{1979PhRvD..20..351Z} and \citet{1980PhRvD..21..891Z}, building upon earlier work on the gravitational-wave emission from rigid, rotating bodies \cite{1970Natur.228..655C,1970ApJ...161L.137C,1973A&A....28..429B}. Since then, there has been development on this topic, particularly in the direction of modelling neutron stars more accurately to understand the associated gravitational-wave signal \cite{2000PhRvD..63b4002C,2001MNRAS.324..811J,2002MNRAS.331..203J,2005CQGra..22.1825V,2020MNRAS.498.1826G}. This work has been partly motivated by some observational evidence for precession, most notably for PSR B1828--11, which spins at a rate of \qty{2.5}{\hertz} with arrival-time residuals of $\approx \qty{500}{\day}$ \cite{2000Natur.406..484S,2001ApJ...556..392L,2006MNRAS.365..653A}, as well as some other sources, including PSR B1642--03 with periodic variations across $\sim \qty{1000}{\day}$ \cite{2001ApJ...552..321S}, potential magnetars in GRB 080602 and GRB 090510 \cite{2020ApJ...892L..34S} and the accreting neutron star in 4U 1820--30 with a precession period of $\sim \qty{1000}{\s}$ \cite{2023MNRAS.523.2663C}. Although based on the present evidence, it would seem that the majority of pulsars do not exhibit long-period precession.

During precession, the star has angular velocity $\Omega^i$ and moves in a cone given by angle $\theta_\text{w}$ about the its angular momentum $J^i$. This angle is the \textit{wobble angle}. If we assume that the star is axisymmetric with respect to its $\bar{z}$-axis, which we ascribe the unit vector $m^i$, in a Cartesian coordinate system, the moment of inertia tensor may be decomposed as \cite{2001MNRAS.324..811J}
\begin{equation}
    I_{i j} = I_0 \, \delta_{i j} + \Delta I \left( m_i m_j - \frac{1}{3} \delta_{i j} \right).
\end{equation}
Connecting with our discussion in Sec.~\ref{sec:Rotating}, this star has principal moments $I_1 = I_2 = I_0 - \Delta I / 3$ and $I_3 = I_0 + 2 \Delta I / 3$, where $\Delta I = I_3 - I_1$. A prolate star has $\Delta I < 0$ and an oblate one has $\Delta I > 0$. The angular momentum is therefore
\begin{equation}
    J_i = I_{i j} \Omega^j = I_1 \Omega_i + \Delta I m_i m_j \Omega^j.
    \label{eq:AngularMomentum}
\end{equation}
This shows that the three vectors $(J^i, \Omega^i, m^i)$ exist in the same plane. The angular momentum is conserved, so we can always adopt a coordinate system in which the other two vectors rotate about $J^i$.

As shown in Fig.~\ref{fig:Precession}, there is another angle $\hat{\theta}$ between $\Omega^i$ and $J^i$. Assuming the star is deformed away from sphericity only slightly such that $|\Delta I| \ll I_0$,
\begin{equation}
    \hat{\theta} \approx \frac{\Delta I}{I_1} \sin \theta_\text{w} \cos \theta_\text{w}.
\end{equation}

It is convenient to define the unit vector $j^i$ in the direction of $J^i$. Therefore, we can construct the angular velocity out of the two unit vectors,
\begin{equation}
    \Omega^i = \dot{\phi} j^i + \dot{\psi} m^i.
\end{equation}
The two degrees of freedom $\dot{\phi}$ and $\dot{\psi}$ describe the star's precession about $J^i$. The symmetry axis $m^i$ rotates about $J^i$ along with angular frequency $\dot{\phi}$. This is typical of rotation. Precession introduces the additional angular frequency $\dot{\psi}$, which characterises the rotation about $m^i$. The three angles $(\phi, \theta_\text{w}, \psi)$ are the familiar Euler angles, which describe the orientation of a body in a coordinate system \cite{LandauLifshitz1976}.

Since we assume that the star undergoes torque-free precession (which will not be so when one considers the back-reaction due to gravitational and electromagnetic radiation), the angular momentum is constant throughout the motion, $J^i = J j^i$, and so Eq.~\eqref{eq:AngularMomentum} provides the two constraints
\begin{subequations}
\begin{gather}
    J = I_1 \dot{\phi}, \label{eq:UsualRotation}\\
    \dot{\psi} = - \frac{\Delta I}{I_3} \dot{\phi} \cos \theta_\text{w}.
\end{gather}
\end{subequations}
Equation~\eqref{eq:UsualRotation} is the usual relationship between the angular momentum of the star and its rotation rate. If $I_1 = I_3$ and we have a rotating, rigid, spherical body, then $\dot{\psi} = 0$ and there is no precession. Note that the precession depends on the wobble angle $\theta_\text{w}$.

The precession impacts the gravitational radiation. Transforming to the inertial frame, we find the energy radiated from the star is
\begin{equation}
    \frac{dE}{dt} = - \frac{2 G}{5 c^5} \Delta I^2 \dot{\phi}^6 \sin^2 \theta_\text{w} (\cos^2 \theta_\text{w} + 16 \sin^2 \theta_\text{w}).
\end{equation}
Here, we see that, in the case of $\theta_\text{w} = 0$, there will be no emission, since the star will be axisymmetric. Whereas, when $\theta_\text{w} = \pi / 2$, we obtain the result~\eqref{eq:RadiateEnergy} from earlier. Calculating the strain is more involved, since it involves the observer's inclination with respect to the star. One finds that a quadrupolar, precessing star will radiate gravitational waves at twice the rotation rate $2 \dot{\phi}$ and at the rotation rate $\dot{\phi}$ of the star \cite{1979PhRvD..20..351Z}. In fact, for small wobble angles, the dominant contribution is at $\dot{\phi}$, in stark contrast with the non-precessing case.

It should, of course, be noted that a realistic neutron star with a magnetic field will radiate electromagnetically, which will influence the evolution of $\dot{\phi}$ and $\dot{\psi}$. Additionally, the back-reaction of the gravitational waves will affect the wobble angle $\theta_\text{w}$. Real neutron stars will exhibit dissipation through, \textit{e.g.}, bulk and shear viscosity that are related to the nuclear reactions in the interior. These dissipative processes will damp the precessional motion.%
\footnote{The damping mechanisms that exist in neutron stars are also important for oscillations. An oscillation mode of the star may be excited through (say) a tidal interaction with a compact companion. As the mode rings, it will be damped by viscosity and gravitational waves. This makes the mode quasi-normal as the energy dissipates. Such damping is particularly relevant for the famous Chandrasekhar-Friedman-Schutz instability \cite[see, \textit{e.g.},][]{2003CQGra..20R.105A}.}
In order for precession to have a meaningful involvement in gravitational-wave emission before it is damped, the star's evolutionary history must be such that its wobble angle is significant. This could involve a number of processes, such as accretion and electromagnetic torques.

\citet{2000NewA....5..243M,2000astro.ph.10044M} claimed to find evidence for an optical pulsar with spin frequency \qty{467.5}{\hertz} in SNR 1987A. The period of precession $2 \pi / \dot{\psi}$ was measured to vary between \qty{935}{\s} and \qty{1430}{\s}. Further, they argued that the observed spin-down was consistent with what one would expect if the precessing star was emitting gravitational waves. This required the wobble angle remaining approximately constant. However, the pulsations vanished after 1996 and were never independently confirmed.

\section{Detection prospects}
\label{sec:Prospects}

Before we conclude, we now consider the opportunities and prospects for gravitational-wave detections of neutron-star mountains. From Eq.~\eqref{eq:StrainAmplitude}, we know that the strain amplitude is exceedingly weak. As we have already mentioned, for a steady signal, we can do better by coherently observing the source for long intervals, building up the effective amplitude. While this is true, completing such an integration is far from trivial and there are several complicating factors.

Suppose we have a rotating neutron star hosting a long-lived, quadrupolar deformation emitting gravitational radiation at a frequency $f$. In the absence of precession, the frequency is related to the star's spin by $f = \Omega / \pi = 2 \nu$. Furthermore, we imagine that in order to measure the waves, we require an observational duration of $T_\text{obs} = \qty{1}{\yr}$. This corresponds to a frequency resolution of $1 / T_\text{obs} \sim \qty{30}{\nano\hertz}$ \cite{2023LRR....26....3R}. However, since the star is radiating gravitational waves, it will gradually spin down according to Eq.~\eqref{eq:SpinDown} and the gravitational-wave frequency will adjust in tandem,
\begin{equation}
    \dot{f} \approx \num{-1.7e-9} \left( \frac{\epsilon}{\num{e-6}} \right)^2 \left( \frac{I_3}{\qty{e45}{\g\cm\squared}} \right) \left( \frac{f}{\qty{1000}{\hertz}} \right)^5 \, \unit{\hertz\per\second}.
    \label{eq:FrequencyDerivative}
\end{equation}
In order for the signal to sit in the frequency bin during the year-long observation, the signal's time derivative $\dot{f}$ must satisfy $1 / T_\text{obs} \gtrsim T_\text{obs} \dot{f}$. This implies that $\dot{f} \lesssim \qty{e-15}{\hertz\per\second}$. Not only are the Doppler modulations due to the Earth's motion much larger than this constraint, but the frequency derivative is also typically much larger. Indeed, Eq.~\eqref{eq:FrequencyDerivative} shows the signal's variation for a source solely influenced by gravitational waves. Of course, these two issues are not particularly severe if the star's spin evolution is well-known and precisely modelled by incorporating gravitational and electromagnetic torques. Then, one can make corrections to keep the signal in the appropriate frequency bin.

Where the problem does get more complicated is when characteristics of the source are not known, or poorly constrained. Whenever there are large uncertainties with the source (which is common), substantial computational costs are incurred as the gravitational-wave search much look over a wide parameter space \cite{2023APh...15302880W,2023LRR....26....3R}. Indeed, some of the most promising sources are accreting pulsars (as discussed in Sec.~\ref{sec:Accretion}). However, we still do not fully understand the accretion torques in these systems and the matter is further complicated by variations in the outburst/quiescence cycles \cite{2021ASSL..461..143P}.

\begin{figure}[h]
    \includegraphics[width=0.6\textwidth]{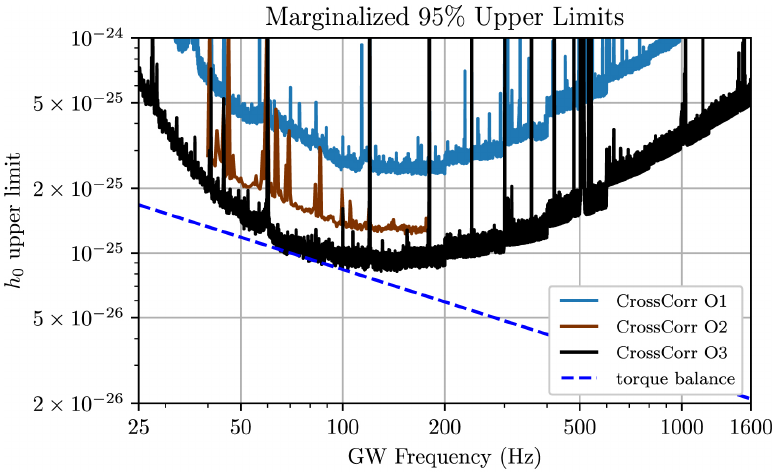}
    \caption{\label{fig:ScoX1}%
    Most recent upper limits on the gravitational-wave strain amplitude $h_0$ with 95\% confidence from directed searches on Scorpius X-1 as a function of gravitational-wave frequency \cite[reproduced from][]{2022ApJ...941L..30A}. The results shown are marginalised over spin inclination. The blue dashed line indicates the limit obtained from torque balance. The upper limits from two previous cross-correlation searches are shown for comparison---\citet[][``CrossCorr O1'']{2017ApJ...847...47A} and \citet[][``CrossCorr O2'']{2021ApJ...906L..14Z}. The most stringent limits lie between \qty{100}{\hertz} and \qty{200}{\hertz} with a sensitivity of $h_0 \approx \num{e-25}$.
    }
\end{figure}

Scorpius X-1 is the most luminous low-mass X-ray binary in the Earth's sky. Under the assumption of torque balance between accretion and gravitational radiation (see Box~\ref{box:Torque}), this makes Scorpius X-1 a natural candidate for gravitational-wave searches \cite{2008MNRAS.389..839W}. The downside is that much about this binary is unknown---in particular, the neutron-star spin frequency is not well constrained \cite{2022MNRAS.509.1745G}. Nevertheless, there have been a number of searches targeting Scorpius X-1, starting with initial LIGO's second observing run \cite{2007PhRvD..76h2001A} and most recently with advanced LIGO's third run \cite{2022ApJ...941L..30A}. To date, no evidence for gravitational waves has been measured, but this has come with upper limits on the gravitational-wave strain amplitude $h_0$. The most recent constraints from  \citet{2022ApJ...941L..30A} are displayed in Fig.~\ref{fig:ScoX1}. Current techniques are now able to probe the torque-balance benchmark at certain frequencies. Searches on Scorpius X-1 would undeniably be aided by observational constraints on its rotation.

For our purposes, Scorpius X-1 serves as an illustrative example, since no continuous gravitational waves have been detected yet from any source. The searches that have been conducted have provided only upper limits on the gravitational waves, but they have also sharpened the sensitivities of methods, which are continually improving \cite[for recent reviews, see][]{2023APh...15302880W,2023LRR....26....3R}. Whether the first detection is imminent or far off remains a mystery. This is in part due to the uncertainties we have discussed in this review, including how large the mountains may be \cite{2015MNRAS.450.2393H}.

It should be noted that we are now entering a regime where gravitational-wave searches are probing astrophysically interesting limits for pulsars. The recent targeted searches in the LIGO-Virgo third observing run have provided the most stringent upper limit to date \cite{2022ApJ...932..133A}. This is for PSR J0711--6830, which has been constrained to have an ellipticity of $\epsilon \lesssim \num{1.7e-8}$. This result, along with $\epsilon \lesssim \num{5.7e-8}$ for PSR J2124--3358, falls below some of the maximum-quadrupole estimates that we discussed in Sec.~\ref{sec:Crust}.

Probing the dense-nuclear structure of neutron stars is a key science objective for the third generation of gravitational-wave observatories, The Einstein Telescope and Cosmic Explorer, which promise significantly enhanced sensitivities \cite{2012CQGra..29l4013S,2021arXiv210909882E}. These instruments are targeting a factor of ten improvement over the advanced detectors in a wide frequency range of \qty{1}{\hertz}--\qty{10}{\kilo\hertz}, which will be useful for searches of rotating neutron stars \cite{2010CQGra..27s4002P,2019BAAS...51g..35R}.

There has been some recent activity focused on population-synthesis studies of the detectability of continuous gravitational-wave emission \cite{2021A&A...649A..92C,2021ApJ...921...89R,2023ApJ...952..123P}, developing upon older work \cite{2005MNRAS.359.1150P,2008PhRvD..78d4031K}. These studies are generally in agreement that we may see gravitational waves from galactic neutron stars with current detectors, but the prospects are far more favourable for third-generation detectors. \cite[See also][for a related study on magnetic deformations.]{2021Galax...9..101S}

In the exciting event that we obtain a confident detection of a continuous gravitational wave, it is natural to ask what could we learn from the signal \cite{2022atcc.book..201J}? From the outset, we should acknowledge that there are other candidates for producing long-lasting gravitational radiation in addition to spinning neutron stars. Many of these are of a speculative nature and include boson clouds around black holes \cite{2018PhRvD..98j3017D}, primordial black holes \cite{2021PDU....3200836M} and the direct interaction of dark photon dark matter with the interferometers \cite{2022PhRvD.105f3030A}. For now, we will assume the source to be a neutron star.

From the theory perspective, there are two basic mechanisms for producing continuous gravitational waves. These are either from a long-lived mountain (the focus of this review) or an excited oscillation mode. Provided that the rotational frequency of the star $\nu$ is known, these two mechanisms are quite easily distinguishable from one another. As we have discussed, a mountain would give rise to gravitational radiation with frequency $f = 2 \nu$. If the neutron star is precessing, deformed by the magnetic field or has a pinned superfluid component, there may also be a signal at $f = \nu$. Whereas an oscillation mode would emit radiation at the frequency of the mode. The most promising oscillation mode from a gravitational-wave perspective is the \textit{r}-mode. For a simple Newtonian model, the $(l, m) = (2, 2)$ \textit{r}-mode has a frequency of $4 \nu / 3$ \cite{1978MNRAS.182..423P}.%
\footnote{The calculation of an \textit{r}-mode becomes more complicated in general relativity. At present, the equations seem to imply a continuous spectrum of frequencies for the relativistic \textit{r}-mode \cite{1998MNRAS.293...49K,2000PhRvD..63b4019L}. It seems unlikely that this result is physical and may be an artefact of our ignorance of the problem. If this is indeed the case, the expectation is that general relativity will provide some correction to the Newtonian mode frequency of $4 \nu / 3$.}

The situation would become significantly more difficult if there were no electromagnetic counterpart and thus no measurement of the spin. This is, in some sense, the worst-case scenario for continuous-wave detection. Suppose the signal was sufficient to determine up to the second time derivative of $f$ (which would require the star to be spinning down relatively rapidly). Then, since we expect $f \propto \nu$, we can determine the braking index $n$ from $f$ and its derivatives (see Box~\ref{box:BrakingIndex}). This would be able to differentiate between a neutron star solely spinning down due to a mountain $n = 5$ and one solely spinning down due to an excited \textit{r}-mode $n = 7$. However, as we have discussed, the neutron stars we observe are rarely so simple.

The central issue in the absence of a measured spin frequency is the following: there exists a degeneracy in the gravitational-wave strain~\eqref{eq:StrainAmplitude} between the three quantities $(d, \epsilon, I_3)$.%
\footnote{This is quite contrary to the situation for gravitationally radiating binaries, where the signal alone is sufficient to calculate the distance to the source \cite{1986Natur.323..310S,2017Natur.551...85A}.}
\citet{2022MNRAS.509.5179S} have discussed in detail how this degeneracy may be broken. Under the assumption that the spin-down is dominated by gravitational radiation, the signal will evolve according to Eq.~\eqref{eq:FrequencyDerivative}. Based on realistic equations of state, the moment of inertia $I_3$ generally varies by only a factor of $2$. This allows an estimate of the ellipticity $\epsilon$ to within a factor of $\sqrt{2}$ and the distance $d$ can be similarly constrained. \citet{2023MNRAS.521.1924S} suggest that gravitational-wave parallax measurements of the source would be useful in further breaking the degeneracy. However, given the ignorance of the star's rotation, one cannot be sure that the gravitational wave is from a mountain; it may just as well be from an oscillation.

It should also be noted that with or without a spin measurement, it will be difficult to distinguish between mountain formation channels. A strong assumption that most analyses make is to assume that the mountain is constant throughout the observation. This should be reasonable for an isolated, well-behaved neutron star, where the deformation is sustained by crustal strains (Sec.~\ref{sec:Crust}) or the magnetic field (Sec.~\ref{sec:Magnetic}), but this will not be the case for more complicated, dynamical sources, like an accreting neutron star (Sec.~\ref{sec:Accretion}). One interesting possibility for a gravitational-wave detection of an accreting pulsar is that the in-principle measurement of cyclotron resonance scattering features in the X-ray spectrum may distinguish between thermal and magnetically confined mountains \cite{2014MNRAS.445.2710P,2015MNRAS.450.2393H}.

\section{Summary}
\label{sec:Summary}

The future looks bright when it comes to gravitational-wave astronomy. Current-generation, ground-based detectors have seen nearly 100 compact-binary coalescences since 2015 and several of these sources include neutron stars. Aside from binaries, we know that non-axisymmetrically deformed, spinning neutron stars will radiate gravitationally, but the expected amplitude of the radiation is very weak. There is a substantial effort ongoing in developing and improving the necessary methods to extract such weak signals from the strain data, along with plans for the construction of third-generation detectors well in development. Therefore, there is good reason to remain optimistic about the possibility of detecting continuous gravitational waves from rotating neutron stars.

In this review, we discussed the status of the theory on neutron-star mountains. We motivated why mountains are of interest from a gravitational-wave perspective and briefly considered the electromagnetic evidence for neutron stars spinning down due to gravitational radiation. The evidence is far from conclusive, but it does illustrate a complementary approach to searching for gravitational waves indirectly---akin to the gravitational waves inferred from the radio observations of the famous Hulse-Taylor binary \cite{1975ApJ...195L..51H,1975ApJ...196L..63W}. We reviewed the mechanisms that would give rise to such deformations on a neutron star. These include (i) elastic strains in the crust that were produced during the star's evolution, (ii) the magnetic field distorting the stellar shape and (iii) accretion producing either a thermal or magnetically confined mountain. All of these mechanisms would produce a gravitational-wave signal at twice the star's spin frequency. Later, we discussed how precession enters the picture and modifies the associated gravitational radiation. There is some observational evidence for precession in neutron stars and a gravitational-wave detection would be particularly interesting, since it would come with an additional frequency component, as compared to the non-precessing case. Finally, we considered the current prospects for detection, in particular, discussing the difficulties that arise in the absence of electromagnetic information about the source and the degeneracy problem. Indeed, the degeneracy that exists in a gravitational-wave signal is quite severe and presents a serious hinderance for inferring aspects about the involved neutron star. Furthermore, without additional information on the neutron star's spin, it will be difficult to distinguish whether the signal comes from a mountain or an oscillation mode, or from an entirely different gravitational-wave source altogether.

From the modelling perspective, there is no shortage of problems to work on. However, it should be emphasised that much of our present understanding comes from calculations of equilibrium neutron stars and we lack sophisticated, evolutionary simulations. Evolutionary computations will be necessary in order to make progress understanding what deformed shapes typical neutron stars take and their ellipticities. Of course, such simulations will inevitably require inputting aspects of physics that are still highly uncertain, such as accretion and magnetic fields. Although speculative, it seems entirely plausible that exploring these issues from the theory may lead to opportunities to break the existing degeneracy problem.

Since the focus of this review has been on long-lived mountains that will produce continuous gravitational waves, we have not discussed in detail transient sources. We mentioned that fall-back accretion following a supernova or merger may give rise to gravitational radiation. In addition to this, there is the possibility that glitches in neutron stars could be associated with the development of a transient asymmetry \cite{2019PhRvD.100f4058K,2020MNRAS.498.3138Y,2023MNRAS.518.4322Y,2023MNRAS.519.5161M}.

This review highlights that there is a wealth of neutron-star physics involved in sourcing mountains. While there are many questions that remain unanswered from the theory front, it is likely that observations will play an important role in providing the answers. With the variety of exotic physics that neutron stars harbour, the prospect of probing their interiors with gravitational waves is quite tantalising indeed.

\begin{acknowledgments}
The author acknowledges support from STFC via grant number ST/V000551/1. The author is grateful for useful discussions and comments from N. Andersson and T. J. Hutchins.
\end{acknowledgments}

\bibliography{bibliography}

\end{document}